\documentclass[prb,superscriptaddress,nobalancelastpage,twocolumn, showpacs,notitlepage]{revtex4-1}
\usepackage[normalem]{ulem}
\usepackage{cancel}
\usepackage{graphicx}
\usepackage{amssymb}
\usepackage{natbib}
\usepackage{dcolumn}
\usepackage{bm}
\usepackage[caption=false]{subfig}
\usepackage{graphicx,psfrag,xspace}
\usepackage{xcolor}
\usepackage{amssymb,amsfonts,amsmath}
\usepackage{setspace}
\usepackage{mathrsfs}
\usepackage{dsfont}
\usepackage{tabularx,booktabs}
\DeclareMathAlphabet{\mathpzc}{OT1}{pzc}{m}{it}
\newcolumntype{Y}{>{\centering\arraybackslash}X}

\def\bra#1{\mathinner{\langle{#1}|}} 
\def\ket#1{\mathinner{|{#1\,}\rangle}} 

\def\omw{\omega_{ \text{\footnotesize MW}}}
\usepackage{comment}

\newcommand{\HH}[1]{\hat{H}_{\text{#1}}}
\def\Hs{\HH{S}}

\def\Hmw{\HH{MW}}
\def\Hhf{\HH{hf}}
\def\rhot{\rho_{\text{tot}}}
\def\rhos{\rho}
\def\rhogibbs{\rho^{\text{\tiny Gibbs}}}

\def\rhol{\rho_{\text{latt}}}
\def\rhor{\rho^{\text{(r)}}}

\def\omw{\omega_{\text{MW}}}
\def\Ut{\hat{U}(t)}
\def\Udt{\hat{U}^\dagger(t)}
\usepackage{empheq}

\usepackage{lipsum}
\usepackage{mathtools}

\newcommand{\titleinfo}{Eigenstate versus Zeeman-based approaches to the solid-effect}

\newcommand{\Tr}{\operatorname{Tr}}

\def\rhostat{\rhos^{\text{stat}}}

\def\pnstat{\pi_n^{\mbox{\tiny stat}}}

\def\psum{\sideset{}{'}\sum}
\def\Leff{L_{\mbox{\tiny eff}}}

\graphicspath{{figures/}}

\def\Ker{\operatorname{Ker}}

\begin{document}

\author{In\'es Rodr\'iguez-Arias}
\affiliation{LPTMS, CNRS, Univ. Paris-Sud, Universit\'e Paris-Saclay, 91405 Orsay, France}
\author{Alberto Rosso}
\affiliation{LPTMS, CNRS, Univ. Paris-Sud, Universit\'e Paris-Saclay, 91405 Orsay, France}
\author{Andrea De Luca}
\affiliation{The Rudolf Peierls Centre for Theoretical Physics, Oxford University, Oxford, OX1 3NP, United Kingdom}

\newcommand{\andrea}[1]{{\color{purple} #1}}
\newcommand{\andreaComment}[1]{\andrea{[\textbf{An: #1}]}}

\title{\titleinfo}
\begin{abstract}

The solid effect is one of the simplest and most effective mechanisms for  Dynamic Nuclear Polarization. It involves the exchange of polarization between one electron and one nuclear spin coupled via the hyperfine interaction. Even for such a small spin system, the theoretical understanding is complicated by the contact with the lattice and the microwave irradiation.  Both being weak, they can be treated within perturbation theory. In this work, we analyze the two most popular perturbation schemes: the Zeeman and the eigenstate-based approaches which differ in the way the hyperfine interaction is treated. For both schemes, we derive from first principles an effective Liouville equation which describes the density matrix of the spin system; we then study numerically the behavior of the nuclear polarization for several values of the hyperfine coupling. In general, we obtain that the Zeeman-based approach underestimates the value of the nuclear polarization. By performing a projection onto the diagonal part of the spin-system density matrix, we are able to understand the origin of the discrepancy, which is due to the presence of parasite leakage transitions appearing whenever the Zeeman basis is employed.
\end{abstract}

\maketitle
\section{Introduction}\label{Intro}

Dynamic nuclear polarization (DNP) is an extremely promising technique to improve the signal-to-noise ratio in magnetic resonance imaging (MRI). DNP was predicted and discovered in the fifties,~\citep{Overhauser,OverhauserExp} and is nowadays a powerful method to enhance the polarization of nuclear spins, both in the solid~\citep{Abragam1982a} and liquid~\citep{LiquidDNP} states. In the DNP setup,\citep{Ardenkjer-Larsen2003,Ardenkjer-Larsen2008} a compound doped with electron radicals, at low temperatures and in the presence of a magnetic field, is irradiated with microwaves at a frequency close to the electron Larmor frequency. After some time, a steady state is reached where polarization has been transferred from the electron to the nuclear spins. The stationary nuclear polarization reaches extremely high values well-beyond those expected at thermal equilibrium in similar conditions.

A large number of mechanisms are responsible for such a polarization transfer and which one is the most relevant among them depends on the temperature, the magnetic field and the radical's concentration.~\citep{Abragam1982a,WenckebachDNP} Two theoretical approaches are employed for understanding DNP: 
i) a macroscopic approach based on the rate equation for the  polarizations of nuclear and electron spins, where the polarization transfer is governed by transition rates phenomenologically obtained;~\citep{WenckebachDNP,ColomboSerra2014,C3CP44667K}
ii) a microscopic approach based on the density matrix formalism that takes into account interactions, lattice and microwave irradiation.~\citep{hovav2010theoretical, karabanov2012quantum,deluca2016} 
The microscopic description of the DNP protocol has the advantage of indicating the most effective mechanisms and the main obstacles for the hyperpolarization of the nuclear spins given the setup conditions. The first step in this direction is to derive an effective equation for the time evolution of the density matrix of the spin system. This task cannot be done exactly, but it requires approximations. In particular, the Zeeman interaction with the magnetic field is much stronger than the coupling with the lattice and the action of the microwaves, which can be treated within perturbation theory. As a result the effective time evolution of the density matrix is obtained from a perturbative expansion up to the second order.\citep{petruccione2002theory}

How to account for hyperfine and dipolar interactions between spins is more controversial. The protocol developed in [\onlinecite{hovav2012,Hovav2013,deluca2015,deluca2016}] is based on the exact eigenstates of the spin Hamiltonian, which includes the dipolar and hyperfine interactions. On the contrary, the protocol developed in [\onlinecite{karabanov2012quantum,karabanov2015dynamic}] treats the interactions perturbatively and is based on the Zeeman eigenstates. The former approach thus requires the exact diagonalization of the interacting spin Hamiltonian, while the latter can be described with a much simpler Zeeman basis. 

An important tool for the microscopical understanding of  DNP are numerical simulations, but with the strong limitation that their complexity grows exponentially with the total number of spins, $N$.~\citep{ARMSTRONGSimulations, hovav2010theoretical, karabanov2012quantum}
In particular, the Liouville scheme, which targets time evolution and steady state of the full spin density matrix, requires the diagonalization of $2^{N} \times 2^{N} $ matrices and one is restricted to at most $N\approx7$ spins. One can reach system sizes of  $N\approx 14$  within the \textit{Hilbert approximation} obtained by projecting the time evolution onto the diagonal elements only and reducing the dynamics to transitions between the $2^N$ eigenstates.\citep{Hovav2013,deluca2015}

In this paper, we derive and compare the time evolution obtained within the eigenstate and the Zeeman-based approaches. For simplicity, we focus on the minimal DNP system: an electron spin hyperpolarizes a nuclear spin via the well-resolved solid effect, which consists in the microwave irradiation of forbidden two-spin transitions, ultimately allowed by hyperfine interactions.~\citep{abragam1978principles,JeffriesSE} In both cases the evolution equations for the density matrix are derived using the Lindblad formalism that can be easily generalized to more complex spin systems. Moreover we show that using the Schrieffer-Wolf perturbation theory~\citep{SchriefferWolf} the transition rates of the Hilbert approximation can be systematically computed.  We prove that for weak hyperfine interactions, both time evolutions provide similar results. On the contrary, for large values of the hyperfine interactions, the Zeeman approach is no longer reliable, as parasite leakage transitions come to play. Nevertheless, the Zeeman approach is very useful to study how the nuclear spin hyperpolarization diffuses in real systems as the transitions rates between Zeeman eigenstates do not require matrix diagonalization and can be performed  for $N$ very large with Monte Carlo methods.~\citep{Karabanov-MonteCarlo}

The paper is organized as follows: In Sec.~\ref{DMt-ev} we derive the general form of the time evolution of the reduced density matrix within the Lindblad formalism. In Sec.~\ref{Hilbert}, we further simplify the time evolution by restricting ourselves to the diagonal part of the reduced density matrix $\rhos$ (the aforementioned Hilbert approximation), and in Sec.~\ref{Results}, we present the numerical results for the steady state obtained for the eigenstate and the Zeeman-based approaches using both the Liouville and the Hilbert scheme.

\section{Derivation of the time evolution equation for the spin density matrix\label{DMt-ev}}
\subsection*{The Hamiltonian}
We consider an electron and a nuclear spin, interacting via hyperfine interactions and weakly coupled to the lattice, which plays the role of a thermal bath at the temperature $\beta^{-1}$). The spins are irradiated by a microwave field resonant at $\omw$. The full Hamiltonian reads:
\begin{equation}
\label{Htot}
\HH{tot}  = \underbrace{ \HH{Z} + \Hhf}_{\Hs} + \Hmw   +  \HH{S-L}  + \HH{L} \;.
\end{equation}

Let us go through all the terms in~\eqref{Htot}, one by one: 
\begin{itemize}
\item The time-independent Hamiltonian $\Hs$ contains the spins' degrees of freedom only. In our two-spins system, it writes as the sum of the Zeeman and hyperfine contributions\cite{hovav2015electron,hovav2010,hovav2012,Karabanov2012}:
\begin{align}
\HH{Z} &= \hbar \omega_e \hat{S}_z + \hbar \omega_n \hat{I}_z \;,\label{Hzeeman}\\
\Hhf &= \hbar  \hat{\vec{S}}\bar{\bar{A}} \hat{\vec{I}}
\label{HhfFull}
\end{align}
where $\hat{S}$/$\hat{I}$ are the electron/nuclear spin operators, $\omega_{e/n}$ their respective Larmor frequencies. $\bar{\bar{A}}$ is the hyperfine interaction matrix, that includes both isotropic and anisotropic contributions.\cite{abragam1978principles,Abragam1982a,WenckebachDNP} 
\item The microwave Hamiltonian $\Hmw$ is time-dependent and reads:
\begin{equation}
\label{MWham}
\Hmw =2\hbar \omega_1 \hat{S}_x \cos(\omw t)\;.
\end{equation}
To avoid dealing with an explicitly time-dependent problem, here we employ the \emph{rotating wave} approximation (RWA),~\cite{RWA_PRL} that will be detailed at the end of this section. It is based on considering a reference frame rotating at frequency $\omw$ and neglecting terms with fast frequencies (i.e. $2\omw$).  With this approximation the Hamiltonian in the rotating frame becomes time-independent. 
\item We write the coupling between the lattice and the spin system in the form:
\begin{equation}
\label{bathcouple}
\HH{S-L}= \sum_{\substack{\alpha=x,y,z}} \left( \lambda_S \hat{S}_\alpha\hat{\phi}_\alpha^S + \lambda_I \hat{I}_\alpha\hat{\phi}_\alpha^I \right)\;,
\end{equation}
where $\hat{\phi}_\alpha^S$ and $\hat{\phi}_\alpha^I$ are the lattice modes that linearly couple to the spin operators. The constants $\lambda_S$ and $\lambda_I$ describe the strength of the coupling with the two spin species. 
\item $\HH{L}$ contains the lattice modes which we assume to be at thermal equilibrium at the temperature $\beta^{-1}$. We will see\cite{petruccione2002theory} that the detailed form of this Hamiltonian is not important for the evolution of the spin system. 
\end{itemize}

\subsection*{Time evolution of the reduced density matrix}
The time evolution\citep{DensityMatrix} for the density matrix $\rhot$ of an isolated system described by the Hamiltonian $\HH{tot}$ is given by the equation:
\begin{equation}
\label{full-tev}
\frac{d \rhot}{dt} = -\frac{i}{\hbar} \left[ \HH{tot},\rhot \right]\;.
\end{equation}
The Hamiltonian given in~\eqref{Htot} encodes all the degrees of freedom of both the spin system and the lattice, so obtaining the exact time evolution from Eq.~\eqref{full-tev} is an impossible task. One then needs to turn to some approximations in order to treat the problem. 
As the couplings between the spins' degrees of freedom and microwave and lattice are much weaker than the Zeeman term ($\lambda_S, \lambda_I \ll \hbar \omega_n$), we can focus on the effective time evolution of the reduced density matrix of the spin system once the lattice degrees of freedom are traced out:
\begin{equation}
\label{redrho}
\rhos = \underset{\text{ lattice}}{\Tr} \rhot\;.
\end{equation} 
A set of approximations needs to be performed in order to obtain the effective time evolution of $\rhos$ from Eq.~\eqref{full-tev}:
\begin{itemize}
\item The approximation of \emph{weak-coupling} between the spin system and the lattice. In practice this allows performing a second-order perturbative expansion in $\lambda_I, \lambda_S$ that leads to an effective time evolution for  the much smaller reduced density matrix $\rhos$ instead of $\rhot$.
\item The \emph{Born-Markov} approximation which supposes that the characteristic time of the lattice is much faster than the spin-lattice relaxation times, $T_{1e}$ and $T_{1n}$. In this limit the lattice remains at thermal equilibrium at temperature $\beta^{-1}$ and its state is not influenced by that of the spins. As a result, the reduced density matrix $\rhos$ at time $t+dt$ only depends on the state at time $t$ instead that from the full history at times $t'<t$.
\item The approximations above do not guarantee that the resulting evolution for $\rhos(t)$ is physical: it should additionally be linear and preserve trace and positivity of $\rhos$. This can be enforced if one also assumes the \emph{secular} approximation, according to which oscillating phases in off-diagonal elements of  $\rhos$ are neglected.\cite{petruccione2002theory} The validity of this approximation relies on the assumption that $T_{2,e},T_{2,n} \ll T_{1,e},T_{1,n}$.
\end{itemize}
These hypothesis lead to a Lindblad formulation of the dynamics of the spin system, the full derivation being detailed in appendix~\ref{Lattice_coup} and in reference [\onlinecite{petruccione2002theory}]
\begin{equation}
\label{time_ev}
\frac{d\rhos}{dt} = -\frac{ i}{\hbar} \left[\HH{Z} + \Hhf + \Hmw,\rhos\right]+ \mathcal{L}[\rhos]\;.
\end{equation}
The commutator in~\eqref{time_ev} corresponds to the standard time evolution of an isolated quantum system, while the last term corresponds to the Lindblad super-operator $\mathcal{L}$; it is responsible for a non-unitary evolution which still acts linearly and preserves trace, hermiticity and positivity of $\rhos$. These requirements strongly constrain its form, and the final result reads (see appendix~\ref{Lattice_coup}):
\begin{align}
\label{lind}
\mathcal{L}[\bullet] &= \sum_{\omega, \hat{O}\in\mathcal{O}} J_{\hat{O}}(\omega)\; \mathcal{L}_{\hat{O}_\omega}[\bullet]
\nonumber\\
&= \sum_{\omega, \hat{O}\in\mathcal{O}} J_{\hat{O}}(\omega) \left( \hat{O}_\omega\bullet \hat{O}_\omega^{\dagger}-\frac{1}{2}\{\hat{O}_\omega^{\dagger} \hat{O}_\omega,\bullet \}\right)\;,
\end{align}
where $J_{\hat{O}}(\omega)$ is the spectral function
\begin{equation}
\label{spectralF}
J_{\hat{O}}(\omega)=\lambda_{\hat{O}}^2\sum_{\alpha=x,y,z}\int_{-\infty}^{\infty} ds\; e^{i\omega s}\underset{\text{ lattice}}{\Tr}\left[\rhol\,\hat{\phi}_\alpha^{\hat{O}}(s) \hat{\phi}_\alpha^{\hat{O}}(0) \right]
\end{equation}
and $\mathcal{O}=\{\hat{S}_x,\hat{S}_y,\hat{S}_z,\hat{I}_x,\hat{I}_y,\hat{I}_z\}$ are the different spin-flip operators linearly coupled to the lattice modes. To understand the subscript in $ \hat{O}_\omega$ note that in this weak-coupling limit the lattice exchanges energy quanta $\hbar \omega$ with the spins by inducing transitions between the well-resolved energy levels of the unperturbed spin Hamiltonian $\HH{0}$. As a result the sum over $\omega$ runs over all its energy gaps and  $\hat{O}_\omega$ is obtained from $\hat{O}$ selecting only the transitions with an energy gap $\omega$, i.e. it reads:
\begin{equation}
\label{spectralOp}
\hat{O}_\omega=\sum_{\substack{n,m/\\ \epsilon_n-\epsilon_m=\hbar\omega}}\ket{m}\bra{m}\hat{O}\ket{n}\bra{n}\;,
\end{equation}
where $\ket{n}$ and $\ket{m}$ are the eigenstates of the Hamiltonian $\HH{0}$.

Consequently, the precise time evolution of the system then depends on which terms in $\HH{tot}$ are considered to be large and are included in $\HH{0}$, determining the spectrum of well-resolved energy levels. The Zeeman term $\HH{Z}$ is the largest contribution while $\Hmw$ and $\HH{S-L}$ are always weak. How to account for hyperfine interactions is more questionable.

In this work, we compare and discuss two possible choices for $\HH{0}$ considered in the literature:
\begin{itemize}
\item[$(i)$] A \emph{Zeeman-based approach}\cite{karabanov2012quantum,karabanov2015dynamic,KarabanovIJC,karabanovMolPhys} for which we consider $\HH{0}=\HH{Z}$ as non-perturbed Hamiltonian and thus the eigenstates are factorized in the Zeeman basis. Here the hyperfine interactions are treated perturbatively at the same level as the microwave irradiation $\Hmw$ and the lattice coupling $\HH{S-L}$.
\item[$(ii)$] An \emph{eigenstate-based approach}\cite{hovav2010theoretical,HovavSE,hovav2012,Hovav2013,deluca2015,deluca2016,FreeFermions} for which the non-perturbed Hamiltonian is $\HH{0} = \Hs$. This approach treats exactly the hyperfine interactions, but requires the exact diagonalization of the interacting spin Hamiltonian, implying in general a drastic restriction of the accessible system sizes.
\end{itemize}

The difference between the two approaches can be seen also in the absence of microwave irradiation. Indeed within the \emph{Born-Markov} approximation, $\rhol$ is assumed to be at thermal equilibrium which translates into the condition
\begin{equation}
\label{DetBalance}
J_{\hat{O}}(\omega)=e^{\hbar\beta\omega}J_{\hat{O}}(-\omega)\;.
\end{equation}
This implies that the rates of the transitions generated by the Lindblad super-operator respect detailed balance at the temperature $\beta^{-1}$ so that 
\begin{equation}
\mathcal{L}[\rhogibbs] = 0\;,\text{ with } \;\rhogibbs = e^{-\beta\HH{0}}/Z
\end{equation}
Therefore, in the eigenstate-based approach, where $\HH{0}=\Hs$, one finds that $\left[\rhogibbs,\Hs\right]=0$, so that the steady state coincides with Gibbs equilibrium. If we consider the Zeeman-based approach, the dynamics is more complex, as $\left[\rhogibbs,\Hs\right]\neq0$: on the one side the lattice tries to thermalize the spins at the Gibbs equilibrium $e^{-\beta\HH{Z}}/Z$ while the hyperfine interactions induce additional parasite transitions, which slightly modify the final stationary state. 

If the system is irradiated by the microwaves, which are continuously injecting energy, no relaxation to thermal equilibrium is expected. In the rest of the section we will detail how to treat a time-dependent Hamiltonian as the one in Eq.~\eqref{MWham} in order to obtain quantum jumps analogous to the ones previously discussed.

\subsection*{The \emph{rotating-wave} approximation (RWA)}
In order to deal with an effective time-independent Hamiltonian instead of the original one in Eq.~\eqref{MWham}, we perform the so-called \emph{rotating wave} approximation (see appendix~\ref{RWA} and reference [\onlinecite{RWA_PRL}]). In practice, we work in a frame that is rotating at the same frequency as the microwave field $\omw$ and define the density matrix in the rotating frame:
\begin{equation}
\rhor (t)= \Ut\rho(t) \,\Udt\;,
\end{equation}
with $\Ut=e^{i \hat{S}_z\omw t/\hbar}$. 
Once this transformation is applied to $\Hs$, it removes the time dependence in the microwave field, but generates rapidly oscillating terms with frequencies $2 \omw$.
Since $\omw \simeq \omega_e$ is much larger than the other energy scales in $\Hs$, we perform the \emph{rotating-wave} approximation, where all such high-frequency terms are neglected. After this approximation, the Hamiltonian of the spins' system $\Hs$ in the rotating frame has become time-independent and
commutes with $\hat{S}_z$ but not with $\hat{I}_z$. In particular, for the hyperfine Hamiltonian in Eq.~\eqref{HhfFull}, we obtain the simplified pseudo-secular form that reads
\begin{equation}
\Hhf =  \hbar B \hat{S}_z \hat{I}_x \;,
\label{Hhf}
\end{equation}
where, $B$ is the hyperfine strength that depends on the distance between the two spins and in this paper will take values in the range of the tens and hundreds of $2\pi$kHz. For the sake of simplicity, in Eq.~\eqref{Hhf}, we have neglected the secular term $ \propto \hat{S}_z \hat{I}_z$, as it only induces a small shift in the effective Zeeman gaps but does not imply any nuclear spin flip, thus being inessential for the solid-effect transitions considered here. 

Thanks to the RWA, Equation~\eqref{time_ev} once rewritten in the rotating frame assumes the form:
\begin{equation}
\small
\label{time_evRW}
\frac{d\rhor}{dt} = -\frac{ i}{\hbar} \left[ \HH{Z}^{\text{(r)}} + \Hhf + \Hmw^{\text{(r)}},\; \rhor \right] + \mathcal{L}[\rhor]\;,
\end{equation}
where: 
\begin{align}
\footnotesize 
\HH{Z}^{\text{(r)}}&=\HH{Z}-\hbar\omw \hat{S}_z = \hbar (\omega_e-\omw) \hat{S}_z - \hbar \omega_n \hat{I}_z\;,
\\
\label{HmwRWA}
\Hmw^{\text{(r)}}&=\Ut \Hmw \Udt \approx \hbar\omega_1 \hat{S}_x\;.
\end{align} 
Note that the Zeeman Hamiltonian in the rotating frame implies a shift of the electron Larmor frequency $\omega_e\to\omega_e-\omw$. 
In deriving \eqref{time_evRW}, we used that, consistently with the RWA, $[\hat{S}_z, \HH{0}] = 0$ both for the Zeeman and the eigenstate-based approaches, and therefore
\begin{equation}
\Ut\mathcal{L}[\rhos]\Udt=\mathcal{L}[\rhor]\;.
\end{equation}
This is coherent with the fact that the lattice brings the system to thermal equilibrium, which is unchanged in the rotating frame, as $\hat U(t) \rhogibbs \hat U^\dag (t) = \rhogibbs$.

While the RWA is accurate in our DNP context, it does not allow to systematically go beyond the approximation in \eqref{HmwRWA}. More accurate treatments can be obtained considering the average Hamiltonian theory (AHT)~\cite{AHT} or the Floquet theory.\cite{Floquet} The former consists on time-averaging the original Hamiltonian by discretizing the time-intervals and the latter, which is analogous to the Bloch theorem but for temporal periodicity, translates to an expansion in higher harmonics multiple of $\omw$. 
As we explained above, the RWA is equivalent to truncating at the lowest harmonic in the Floquet theory and for simplicity, here we will restrict to doing so. Interestingly, if one considers the second harmonics and applies AHT, a Bloch-Seigert~\cite{Bloch-Siegert} type of shift (extra renormalization for the electron Larmor frequency) would be recovered.

At this point, one can exactly compute the time evolution in Eq.~\eqref{time_evRW}; more directly we can obtain the stationary state $\rhostat$ by setting $d\rhor/dt = 0$ and solving the resulting linear system. This approach has nevertheless an important drawback: the complexity of the problem grows extremely fast, as the number of components of the density matrix $\rhor$ equals $2^{2(N_e + N_n)}$. This can be straightforwardly done for the $2$--spins example ($N_e = N_n = 1$) and the numerical results are shown in Sec.~\ref{Results}. However, as soon as one wants to consider larger systems, this fast exponential growth makes any numerical treatment prohibitive. For this reason, one considers the Hilbert approximation that we detail in the following section.

\section{Hilbert approach: towards a semi-classical master equation\label{Hilbert} }

\begin{table}
\centering
\def\arraystretch{1.1}
\begin{tabularx}{0.5\textwidth}{*{4}{|Y}|}
\hline
Zeeman  & Zeeman  & Exact  & Exact  \\
eigenstates&energies&eigenstates&energies\\
$\{\ket{n}\}$&$\{\epsilon_n\}$&$\{\ket{\tilde{n}}\}$&$\{\tilde{\epsilon}_n\}$\\
\hline 
\end{tabularx}
\def\arraystretch{1.9}
\begin{tabularx}{0.5\textwidth}{*{4}{|Y}|}
\hline
$\ket{0}=\ket{\uparrow_e\;\downarrow_n}$ & $\frac{\hbar(\omega_e+\omega_n)}{2}$ & $\ket{\tilde{0}}=\ket{\uparrow_e \tilde\beta_n}$ & $\frac{\hbar(\omega_e+\Omega_n)}{2}$ \\
\hline
$\ket{1}=\ket{\uparrow_e\;\uparrow_n}$ & $\frac{\hbar(\omega_e-\omega_n)}{2}$ & $\ket{\tilde{1}}=\ket{\uparrow_e\tilde\alpha_n}$ & $\frac{\hbar(\omega_e-\Omega_n)}{2}$ \\
\hline
$\ket{2}=\ket{\downarrow_e\;\downarrow_n}$ & $-\frac{\hbar(\omega_e-\omega_n)}{2}$ & $\ket{\tilde{2}}=\ket{\downarrow_e\tilde\beta^*_n}$ & $-\frac{\hbar(\omega_e-\Omega_n)}{2}$ \\
\hline
$\ket{3}=\ket{\downarrow_e\;\uparrow_n}$ & $-\frac{\hbar(\omega_e+\omega_n)}{2}$ & $\ket{\tilde{3}}=\ket{\downarrow_e\tilde\alpha^*_n}$ & $-\frac{\hbar(\omega_e+\Omega_n)}{2}$ \\
\hline
\end{tabularx}
\caption{Spectrum of the system in the Zeeman and eigenstate-based approaches. }
\label{EigenTable}
\end{table}

\begin{figure*}
\includegraphics[width=0.49\textwidth]{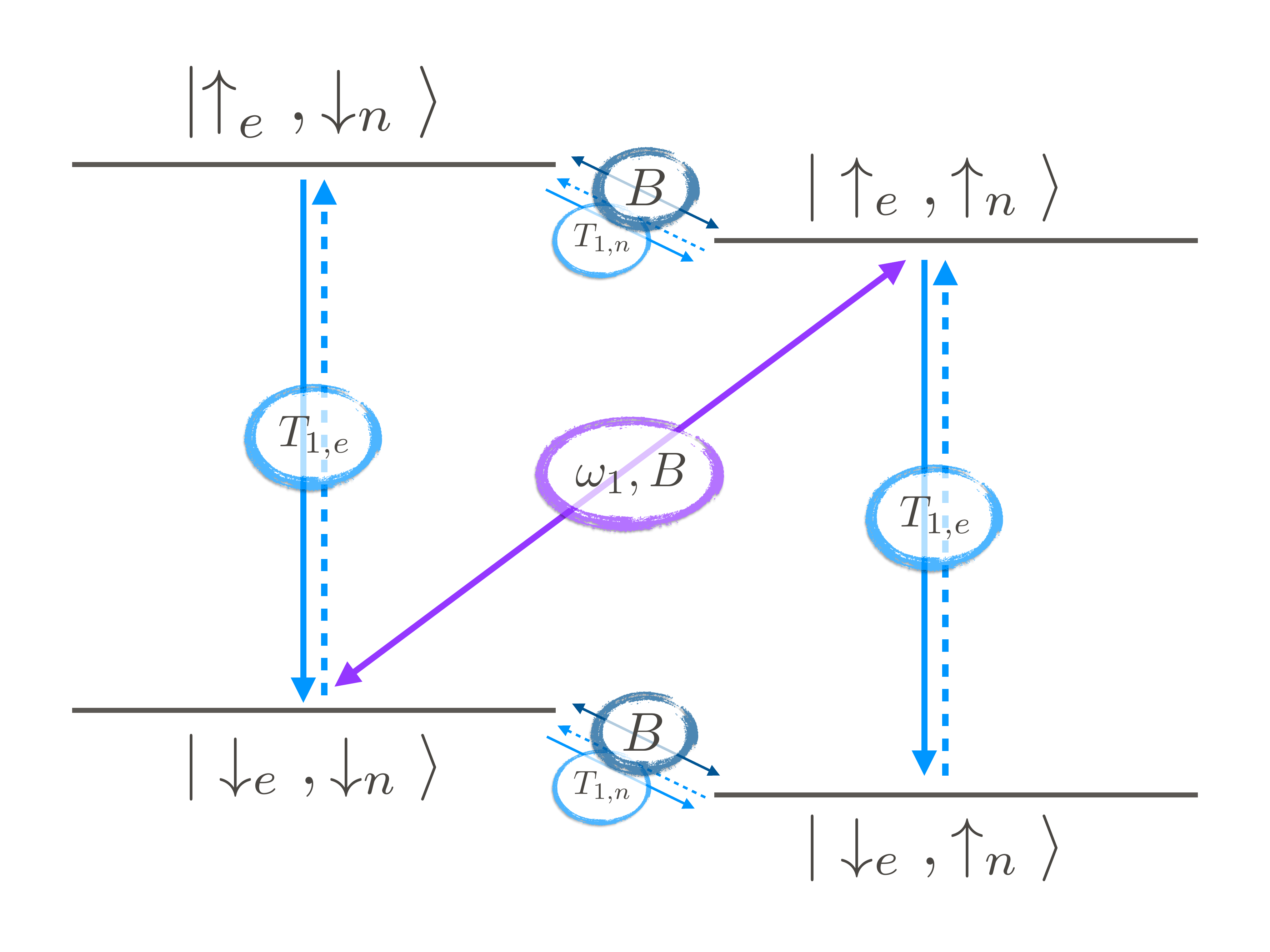}
\includegraphics[width=0.49\textwidth]{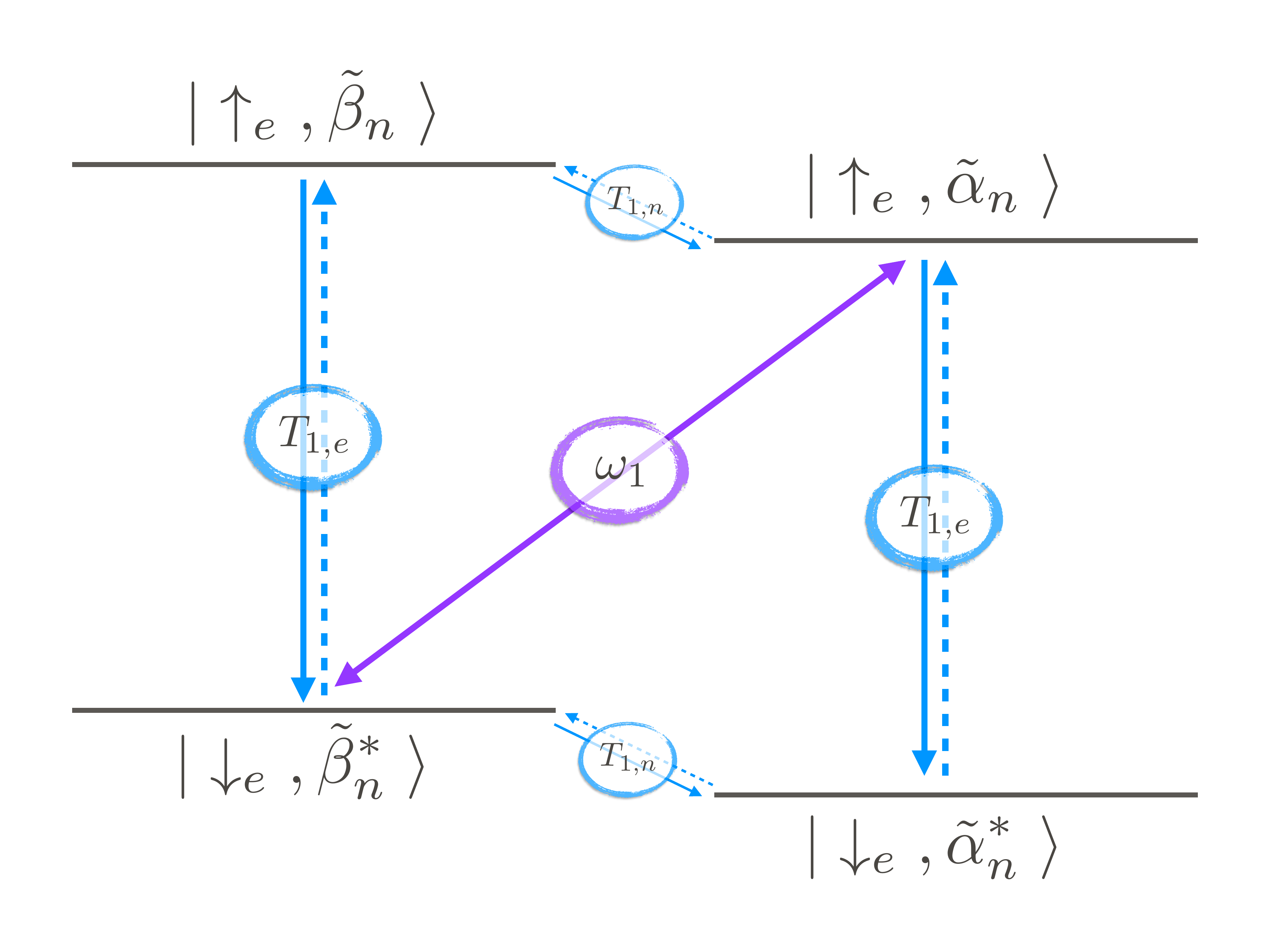}
\caption{(color online) sketch of the possible transitions between eigenstates for the Zeeman-based approach (left) and for the eigenstate-based approach (right). Note that in the Zeeman-based approach, hyperfine interactions ($B$) induce nuclear spin-flip transitions that we dub leakage and that are absent in the eigenstate-based approach. The scheme corresponds to a microwave irradiation of the double-quantum transition (i.e. $\ket{\uparrow_e\;,\uparrow_n}\rightleftharpoons\ket{\downarrow_e\;,\downarrow_n}$), namely $\omw\approx\omega_e-\omega_n$ (or $\omega_e-\Omega_n$). In this limit, we can neglect other microwave-induced transitions (zero-quantum, single-quantum...)}
\label{transitions}
\end{figure*}

The Hilbert approximation consists on projecting the dynamics of the density matrix $\rhos$ to its diagonal components $\rho_{nn}\equiv\pi_n$, in the basis which diagonalizes $\HH{0}$. Indeed, the commutator in \eqref{time_evRW} induces oscillations for the off-diagonal terms
\begin{equation}
\rhos_{nm}(t)\sim \rhos_{nm}(0)e^{i(\epsilon_n-\epsilon_m)t/\hbar}\;,
\end{equation}
which are then exponentially suppressed by the Lindbladian term on time scales $T_{2,e}$ and $T_{2,n}$ (respectively if $\ket{n}$ and $\ket{m}$ differ for an electronic or a nuclear transition). In this limit, the off-diagonal elements are always small and the state of the system can be described by the occupation probability $\pi_n$ of each eigenstate $\ket{n}$. Lattice, microwaves and perturbative terms of the Hamiltonian (if any) induce slow transitions between pairs of those eigenstates. In practice, one wants to obtain a master equation for the occupation probabilities $\pi_n$ and determine the transition rates $W_{\ket{n}\to\ket{m}}$ between eigenstates. Once those rates are determined, one can define the transition matrix $w$:
\begin{equation}
\label{rateMatrix}
w_{nm}=\begin{cases}
\quad \quad  \quad W_{\ket{m}\to\ket{n}}& \text{if }n\neq m \;,
\\
-\sum_{n'}W_{\ket{n}\to\ket{n'}}& \text{if } n=m\;.
\end{cases}
\end{equation}
The stationary state for the occupation probabilities $\vec{\pi}^{\mbox{\tiny stat}}$ is the eigenvector of the matrix $w$ with eigenvalue~$0$, i.e.
\begin{equation}
\label{PIStat}
w\vec{\pi}^{\mbox{\tiny stat}}=0\;.
\end{equation}
The transition rates for the Zeeman and eigenstate-based approaches are in general different. Here we compute them for the two-spins system. As a starting point, the eigenstates and energy levels are given in Table~\ref{EigenTable}. Note that the Zeeman eigenstates 
are completely polarized along the $z$-direction while in the exact eigenstates-based approach, for the nuclear spin, there is some mixing ($\ket{\tilde\alpha_n}$ or $\ket{\tilde\beta_n}$) induced by the hyperfine interactions:
\begin{align}
\label{StatesNucl}
\ket{\tilde\alpha_n} = \cos\varphi\ket{\uparrow_n} - \sin\varphi\ket{\downarrow_n} ,\quad \ket{\tilde\alpha^*_n}=2\hat{I}_z\ket{\tilde\alpha_n},\nonumber\\
\ket{\tilde\beta_n} = \sin\varphi\ket{\uparrow_n} + \cos\varphi\ket{\downarrow_n} ,\quad \ket{\tilde\beta^*_n}=2\hat{I}_z\ket{\tilde\beta_n},\nonumber\\
\end{align}
where $\tan\varphi=\frac{B/2}{\omega_n+\Omega_n}$. Here, we introduced $\Omega_n$ as a shift of the nuclear Larmor frequency:
\begin{equation}
\omega_n\to \Omega_n=\frac{\tilde{\epsilon}_\beta-\tilde{\epsilon}_\alpha}{\hbar}=\omega_n\sqrt{1+\left(\frac{B}{2\omega_n}\right)^2} \;.
\end{equation}
In Fig.~\ref{transitions} we show the possible transitions when the microwaves irradiate at the frequency of the double-quantum transition (i.e. $\ket{\uparrow_e\;,\uparrow_n}\rightleftharpoons\ket{\downarrow_e\;,\downarrow_n}$). 

To derive the expression of the transition rates it is useful to write the Hamiltonian in terms of the non-perturbative term $\HH{0}$ and the perturbation $\hat{V} = \HH{tot} - \HH{0}$. The lattice correlation functions $J_{\hat{O}}(\omega)$ are supposed to be at thermal equilibrium at the temperature $\beta^{-1}$ (see Eq.~\eqref{DetBalance}); in order to specify its structure, we distinguish fast dephasing process at $\omega \simeq 0$ on the scale of $T_{2e}$ and $T_{2n}$ from relaxation decaying process $\omega \neq 0$,
on the scale $T_{1e}, T_{1n}$.  In particular, we set
\begin{equation}
J_{\hat{O}^e}(\omega\neq 0)=\frac{h(\omega)}{T_{1e}} \;, \quad
J_{\hat{O}^n}(\omega\neq 0)=\frac{h(\omega)}{T_{1n}} \;,
\end{equation}
for $\hat O^e = \hat{S}_{x,y}$ ( $O^n= \hat{I}_{x,y}$), with $h(\omega)=\frac{1}{e^{-\beta\omega}+1}$ , and 
\begin{align}
J_{\hat{S}_z}(0) (S_{z,nn}-S_{z,mm})^2\approx \frac{1}{T_{2e}}\;,\nonumber\\
 J_{\hat{I}_z}(0) (I_{z,nn}-I_{z,mm})^2\approx \frac{ 1}{T_{2n}}\;,
\end{align}
for $n\neq m$.

The full dynamics in~\eqref{time_evRW} can now be decomposed in fast contributions with rates of the order $T_{2e}^{-1}, T_{2n}^{-1}$, and slow ones with rates of the order $T_{1e}^{-1}, T_{1n}^{-1}$. Explicitly we have:\footnote{In equation~\eqref{dynamics} we have neglected the two slow contributions of the lattice: $-\rho_{nm} \left(J_{\hat{O}}(\omega_{km}) |O_{km}|^2 + J_{\hat{O}}(\omega_{nk}) |O_{nk}|^2 \right)$}
\begin{widetext}
\begin{align}
\label{dynamics}
\dot{\rho}_{nn} =\sum_{\substack{\hat{O}\in\mathcal{O}\\k\neq n}}& 
\underbrace{|\bra{n}\hat{O}\ket{m}|^2 \left[J_{\hat{O}}(\omega_{nk})\rho_{kk}- J_{\hat{O}}(\omega_{kn})\rho_{nn}\right]
-\frac{i}{\hbar}  \left(\hat{V}_{nk}\rho_{kn} -\hat{V}_{kn}\rho_{nk}\right)}_{\text{slow, }L_1}\;,
\\ 
\dot{\rho}_{nm} =\sum_{\substack{\hat{O}\in\mathcal{O}\\k\neq n,m}}& \underbrace{\left[-\frac{i}{\hbar} (\epsilon_n^{(r)}-\epsilon_m^{(r)})-J_{\hat{O}}(0)(O_{nn}-O_{mm})^2\right] \rho_{nm} }_{\text{fast, }L_0}\;
\underbrace{-\frac{i}{\hbar}  \left(\hat{V}_{nk}\rho_{km} -\hat{V}_{km}\rho_{nk}\right)}_{\text{slow, }L_1}\;,
\end{align}
\end{widetext}
where we define $\hbar\omega_{kn}=\epsilon_k-\epsilon_n$. Note that in $\omega_{kn}$, the energy difference is taken in the lab frame, while the term $\epsilon_n^{(r)}-\epsilon_m^{(r)}$ is in the rotating frame.  
More compactly, we can write
\begin{equation}
\dot{\rho}=L\rho=(L_0+ L_1)\rho\;,
\end{equation}
where $L_0$ accounts for the fast dynamics ($T_{2,e}, T_{2,n}$) while $L_1$ accounts for the slow dynamics ($T_{1,e}, T_{1,n}$ and the perturbative terms in $\hat{V}$). Note that $L_0$ 
is a super-operator which preserves the diagonal part of $\rhos$, i.e.
$L_0[ \ket{n} \bra{n}] = 0$; it has therefore a degenerate subspace corresponding to the projectors on the diagonal entries of the density matrix $\rho_{nn}=\pi_n$, with eigenvalue $0$. 

We are interested in an effective dynamics restricted to the diagonal entries of $\rhos$. As this transitions are induced by the small perturbation $L_1$, we treat the problem perturbatively. We have two contributions: 
\begin{itemize}
\item[$i)$] A dissipative part that acts only on the subspace of the diagonal elements of the density matrix. Thus these lattice-induced transitions come naturally at the first order and read
\begin{subequations}
\label{rates_T1}
\begin{align}
W_{\ket{n}\to\ket{m}}^{T_{1e}}=\frac{h(\omega_{nm})}{T_{1e}} |\bra{n}\hat{O}_e\ket{m}|^2\;,
\\
W_{\ket{n}\to\ket{m}}^{T_{1n}}=\frac{h(\omega_{nm})}{T_{1n}} |\bra{n}\hat{O}_n\ket{m}|^2\;.
\end{align}
\end{subequations}
\item[$ii)$] A part containing the perturbative terms of the Hamiltonian, $\hat{V}$. These terms are responsible for the transition rates which connect different diagonal elements of the density matrix and can be used to build the transition matrix in Eq.~\eqref{rateMatrix}. To do so, we have implemented a perturbation theory based on the Schrieffer-Wolf transformation~\cite{SchriefferWolf} (see appendix~\ref{SW_PT}). In this procedure, we keep only the lowest order term that gives a non-zero contribution between two given eigenstates. 
\end{itemize}

At the second order in $L_1$, one obtains the following transition:
\begin{equation}
\label{2ndOrderRate}
W_{\ket{ n}\to\ket{ m}}^{(2)}= \frac{2|\bra{ n}\hat{V}\ket{ m}|^2}{1+T_{2}^2(\epsilon_n^{(r)}-\epsilon_m^{(r)})^2} \;.
\end{equation}
where $T_{2} = T_2^e$ if the states $\ket{n}$ and $\ket{m}$ differ for an electron spin flip and $T_{2} = T_2^n$ otherwise. The term in Eq.~\eqref{2ndOrderRate} is responsible for the microwave-induced transitions that read:
\begin{equation}
\label{MWrate}
W^{\text{MW}}_{\ket{ n}\to\ket{ m}}=\frac{2 \omega_1^2 T_{2e} |\bra{ n}\hat{S}_x\ket{ m}|^2}{1 + 
 T_{2e}^2 (| \epsilon_n-\epsilon_m|  - \omw)^2}\;.
\end{equation}
In the Zeeman approach, the single-quantum transitions ($\ket{\uparrow_e\;,\downarrow_n} \rightleftharpoons \ket{\downarrow_e\;,\downarrow_n}$ and $\ket{\uparrow_e\;,\uparrow_n} \rightleftharpoons \ket{\downarrow_e\;,\uparrow_n}$) are allowed by this term. Similarly, for the eigenstate-based approach, the transitions  ($\ket{\uparrow_e\;,\tilde{\alpha}_n} \rightleftharpoons \ket{\downarrow_e\;,\tilde{\alpha}^*_n}$ and $\ket{\uparrow_e\;,\tilde{\beta}_n} \rightleftharpoons \ket{\downarrow_e\;,\tilde{\beta}_n^*}$) are induced by this term.

On the contrary, the solid-effect transitions (zero quantum $\ket{\uparrow_e\;,\tilde{\beta}_n} \rightleftharpoons \ket{\downarrow_e\;,\tilde{\alpha}^*_n}$ and double quantum $\ket{\uparrow_e\;,\tilde{\alpha}_n}\rightleftharpoons\ket{\downarrow_e\;,\tilde{\beta}^*_n}$) are only allowed in the eigenstate-based approach. Indeed the numerator has a non-vanishing contribution thank to the mixing of nuclear states in Eq.~\eqref{StatesNucl}. As a result, in the eigenstate-based approach we can restrict ourselves to the second order in the perturbation theory without need to seek into higher-order transitions.

Contrarily, in the Zeeman approach the numerator of Eq.~\eqref{MWrate} vanishes for the zero-quantum and double-quantum transitions. Obtaining the solid effect transition rates in this approach is a tough work and one must go  up to the fourth order in the perturbation theory. The details are given in appendix~\ref{SW_PT}. The final result for this transition requires the joint action of microwave irradiation and hyperfine interactions. It then reads:
\begin{align}
W^{\text{ZQ/DQ}}=
\frac{A^2T_{2e}\omega_1^2}{16\omega_n^2} \bigg(
\frac{1}{1+T_{2e}^2(\omega_e-\omw)^2}+& \nonumber \\ \frac{2}{1+T_{2e}^2(\omega_e\mp\omega_n- \omw)^2}
&\bigg)\;.
\end{align}

An additional consequence of this Zeeman approach is that the hyperfine interactions also induce a transition between different nuclear states. This comes straightforwardly from the second-order formula in Eq.~\eqref{2ndOrderRate}. The transition is dubbed leakage and its rate reads:
\begin{equation}
\label{rateLeak}
W^{\text{B}}_{\ket{ n}\to\ket{ m}}=\frac{ 2B^2 T_{2e} |\bra{ n}\hat{S}_z\hat{I}_x\ket{ m}|^2}{1 + 
 T_{2e}^2 \omega_n^2}\;.
\end{equation}

To sum up, all transition rates between couples of eigenstates are given in appendix~\ref{RateTr}.

\section{Results and discussion}\label{Results}
\begin{figure*}
\includegraphics[width=0.49\textwidth]{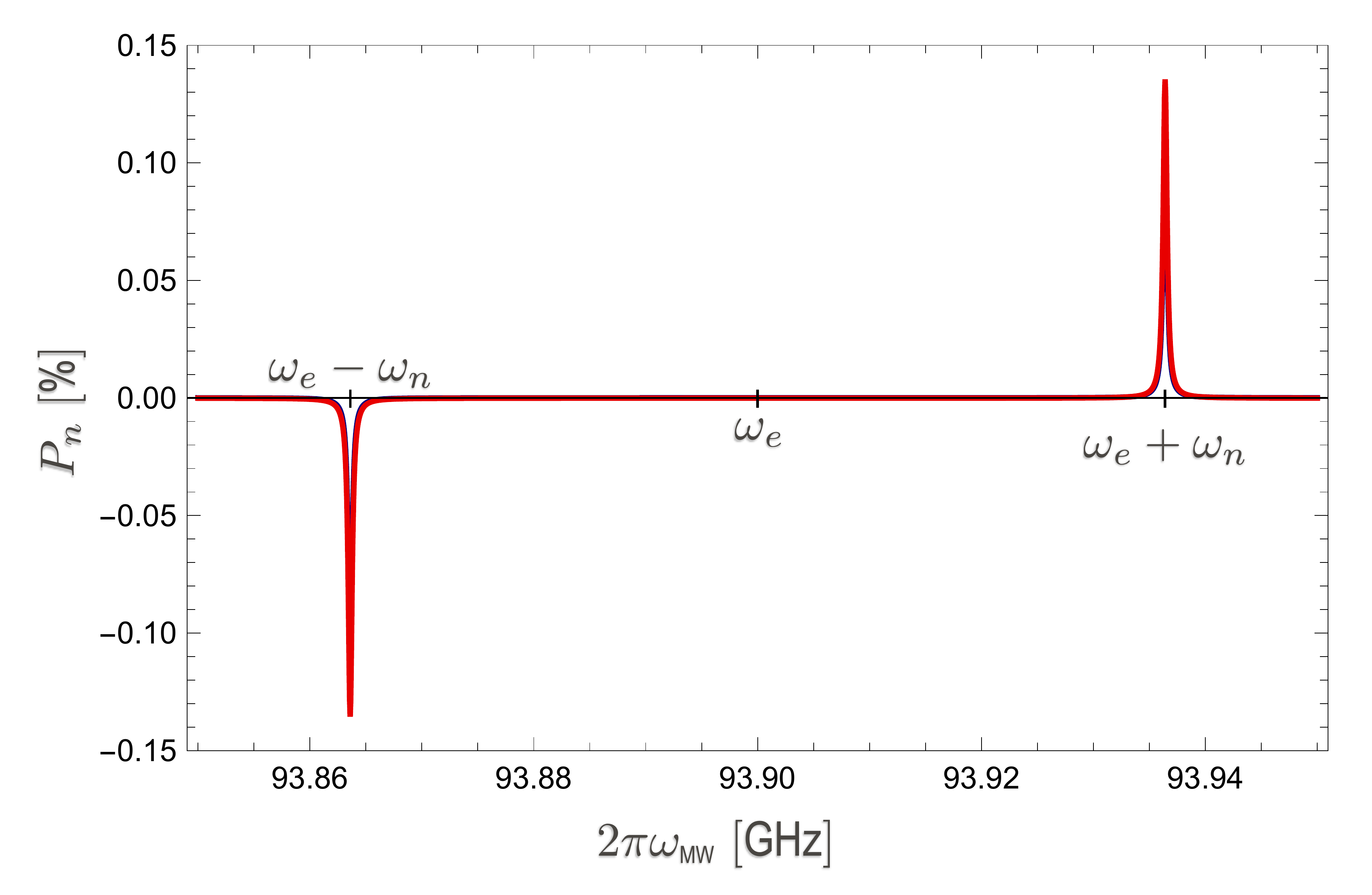}
\includegraphics[width=0.49\textwidth]{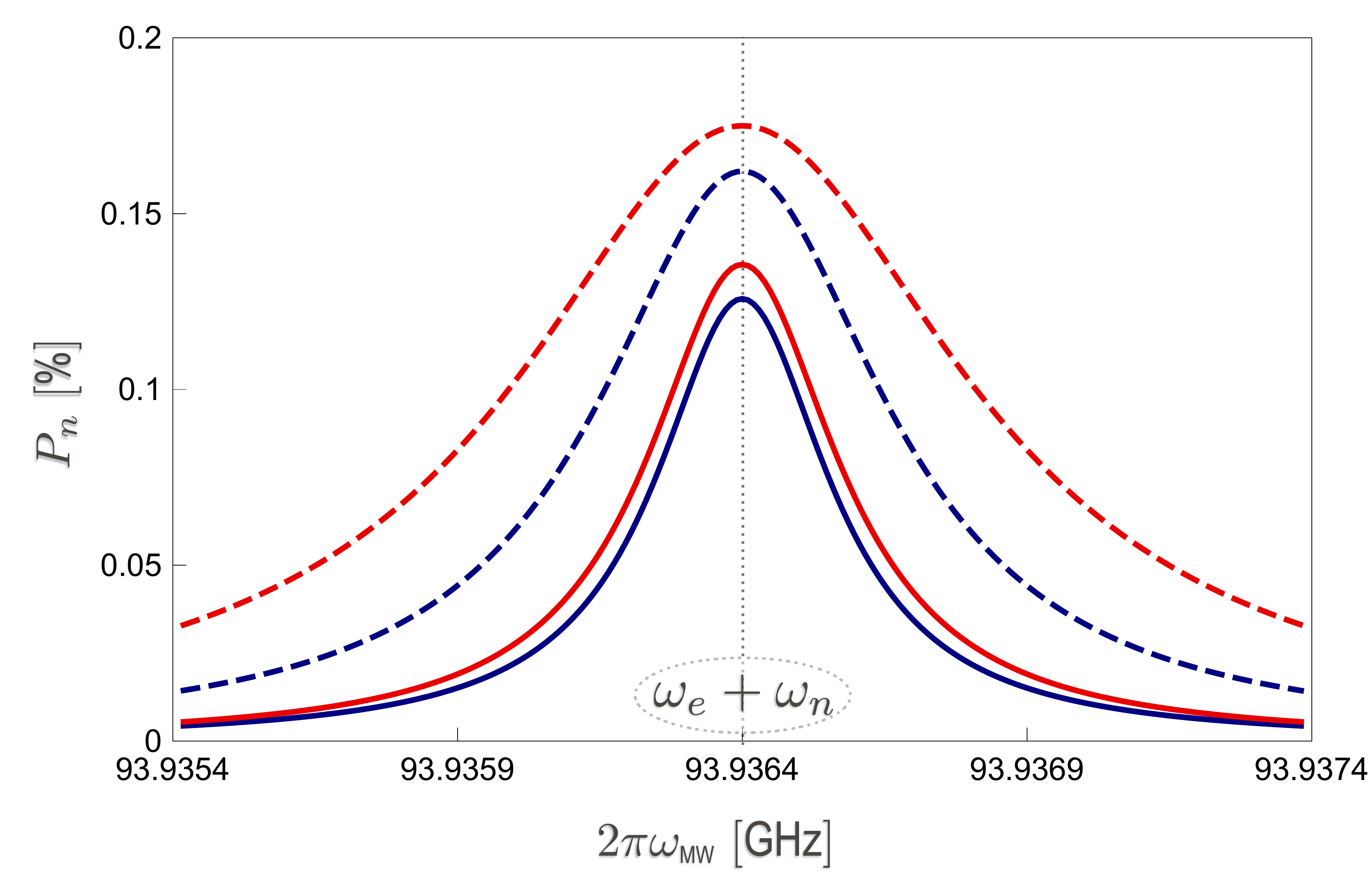}
\caption{(color online) Steady state DNP profile within the Liouville formalism in the Zeeman (blue) and eigenstate-based (red) approaches. (left) Full DNP profile for a hyperfine strength of $B = 40\times2\pi$kHz. We observe the two solid-effect resonances at $\omw\approx\omega_e\pm\omega_n$ corresponding to the double-quantum and zero-quantum transitions. (right) Zoom of the DNP profile around the zero-quantum transition frequency $\omw\approx\omega_e+\omega_n$. The solid line corresponds to the hyperfine strength of $B = 40\times2\pi$kHz while the dashed line shows $B = 160\times2\pi$kHz.}
\label{LiouvillePlot}
\end{figure*}

Following the formalisms introduced in sections~\ref{DMt-ev} and~\ref{Hilbert}, we have computed 
(i) the steady state density matrix $\rhos^{\text{stat}}$ (Liouville scheme) and 
(ii) the occupation probabilities,  $\pnstat$, of the eigenstates of $\hat H_0$ (Hilbert approximation). The nuclear polarization along the $z$ axis reads: 
\begin{equation}
\label{nuclPol}
P_n=  \Tr\left(\rhos^{\text{stat}}  \hat{I}_z\right) \equiv\sum_n \pnstat \bra{n} \hat{I}_z \ket{n}\;.
\end{equation}
The first definition applies for the Liouville scheme while the second one is used within the Hilbert approximation. To compute $\rhos^{\text{stat}}$ we have used Eq.~\eqref{time_evRW} and to compute $\pnstat$, Eq.~\eqref{PIStat} has been employed.

The numerical parameters are given in table~\ref{tableParam} and are chosen to be realistic for an experiment involving $^{13}C$, but for which we vary some physical parameters in order to obtain richer physics.

We now present the study of the DNP profile (i.e. the nuclear polarization as a function of the frequency of irradiation of the microwaves). In particular we will focus on the zero-quantum transition to illustrate our results, but they are applicable all over the spectrum. 

\begin{table}
\begin{center}
\def\arraystretch{1.1}
\begin{tabularx}{0.47\textwidth}{*{4}{|Y}|}
\multicolumn{4}{c}{Electron parameters}\\
\hline
  $T_{1e}$ (s) &  $T_{2e}$ (s) &  $\omega_e$ ($2\pi$GHz) & $\omega_1$ ($2\pi$MHz)   \\
  \hline
  $10^{-3}$ & $10^{-5}$ & $93.9$&  $0.1$    \\
 \hline 
 \multicolumn{4}{c}{ }\\
 \multicolumn{4}{c}{Nuclear parameters}\\
 \hline 
  $T_{1n}$ (s) &  $T_{2n}$ (s) & $\omega_n$ ($2\pi$MHz) & $B$ ($2\pi$kHz)  \\
  \hline 
  $100$ & $5\times 10^{-3}$ & $-36.4$ ($^{13}C$) & $40,160,320$  \\
 \hline  
\end{tabularx}\\
\caption{Microscopic parameters modeling the $1-$electron and $1-$nuclear spin system in a magnetic field of $H=3.3$~T and in contact with lattice at a temperature $\beta^{-1}=12$ K. We have chosen different values for the hyperfine interactions in order to check its action on the relaxation basis.}
\label{tableParam}
\end{center}
\end{table}

\subsection{An exact treatment: Zeeman vs. eigenstate-based approaches within the Liouville formalism}
The full DNP profile obtained from the exact Liouville treatment is shown in Fig.~\ref{LiouvillePlot} (left). 
The resonances corresponding to the zero-quantum and double-quantum transitions at microwave frequencies of $\omw\approx\omega_e\pm\omega_n$ are clearly recognizable. Our interest lies on how the strength of the hyperfine interactions affects the results, and which basis is the most accurate choice given the scenario. It is natural to assume that the Zeeman approach requires the hyperfine strength $B$ to be weak, as it treats such an interaction as a small perturbation.

In Fig.~\ref{LiouvillePlot} (right), we show a zoom of the DNP profile around the zero-quantum transition: red and blue lines correspond respectively to the Zeeman and eigenstate-based approaches. Two values of the hyperfine interactions are considered: in solid line a modest hyperfine strength $B = 40 \times2\pi$kHz is shown, while the dashed lines show the case of $B = 160 \times2\pi$kHz. The width of the peak is proportional to $B$, and we also note that increasing the interaction strength the two eigenstate approaches are significantly more different.

\subsection{Zeeman vs. eigenstate-based approaches in the Hilbert formalism}
\begin{figure*}[ht!]
\includegraphics[width=0.49\textwidth]{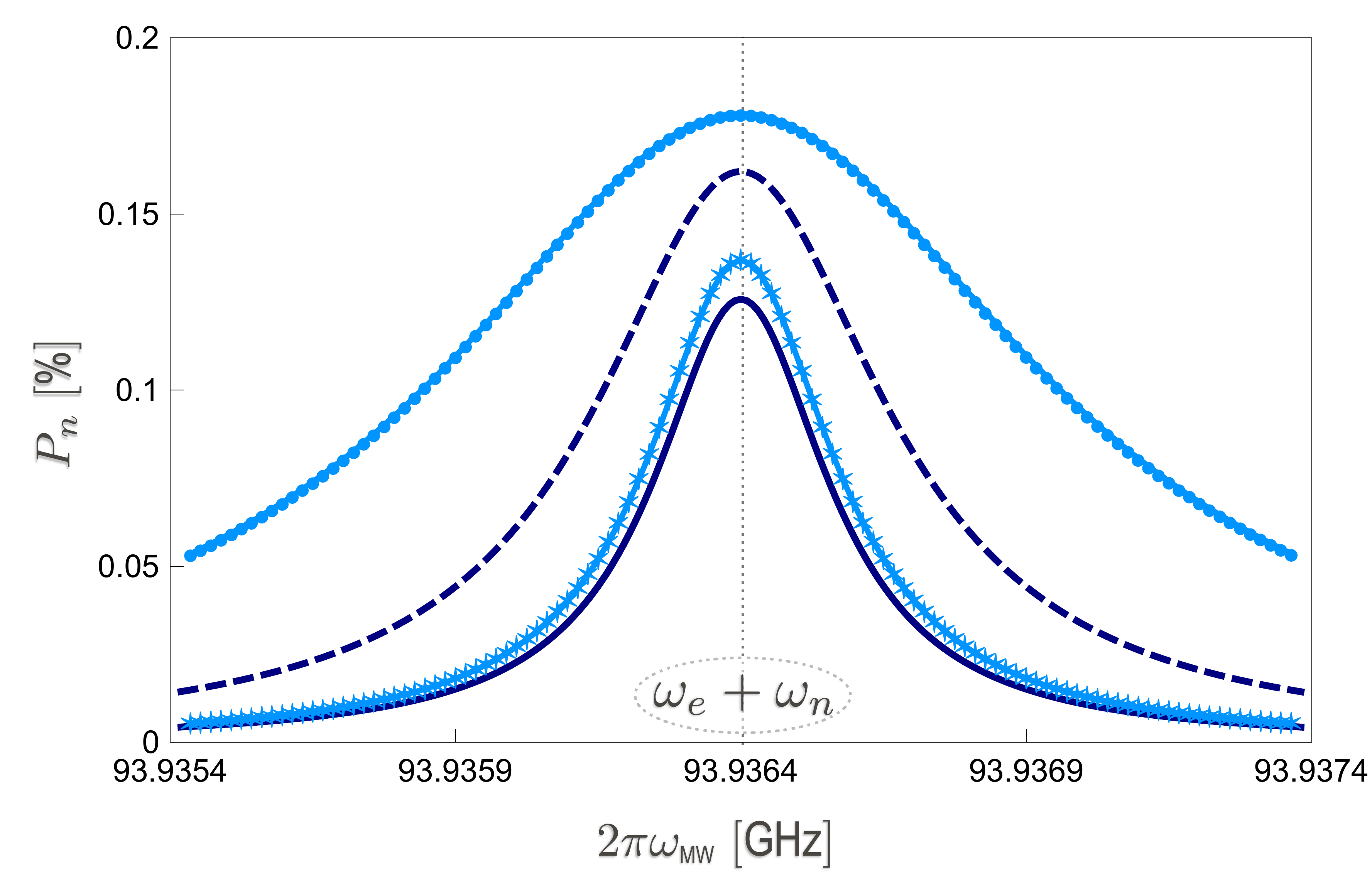}
\includegraphics[width=0.49\textwidth]{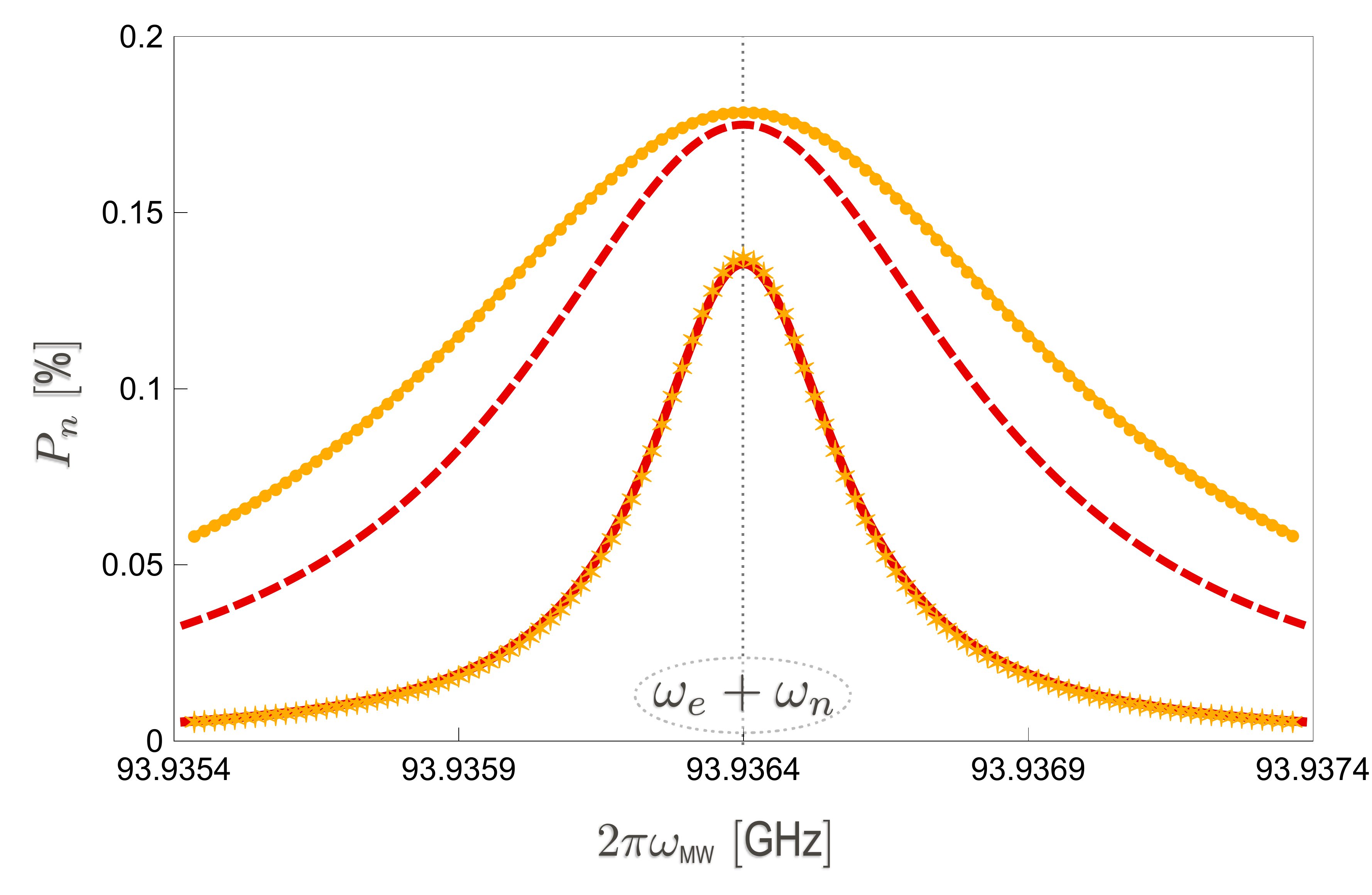}
\caption{(color online) DNP profile around the zero-quantum transition. (left) An exact Liouville treatment in the Zeeman approach is plotted in navy. Different values of the hyperfine interaction are considered: $B = 40 \times2\pi$kHz (solid lines) and $B = 160\times2\pi$kHz (dashed lines). In light blue we show the Hilbert approach in stars/circles for the same two values of the hyperfine interaction.
(right) An analogous plot to the left one in the eigenstate-based approach. The red lines show the exact Liouville formalism and the yellow ones show the Hilbert one.
}
\label{HilbertPlot}
\end{figure*}

\begin{center}
\begin{figure}[!ht]
\centering
\includegraphics[width=0.49\textwidth]{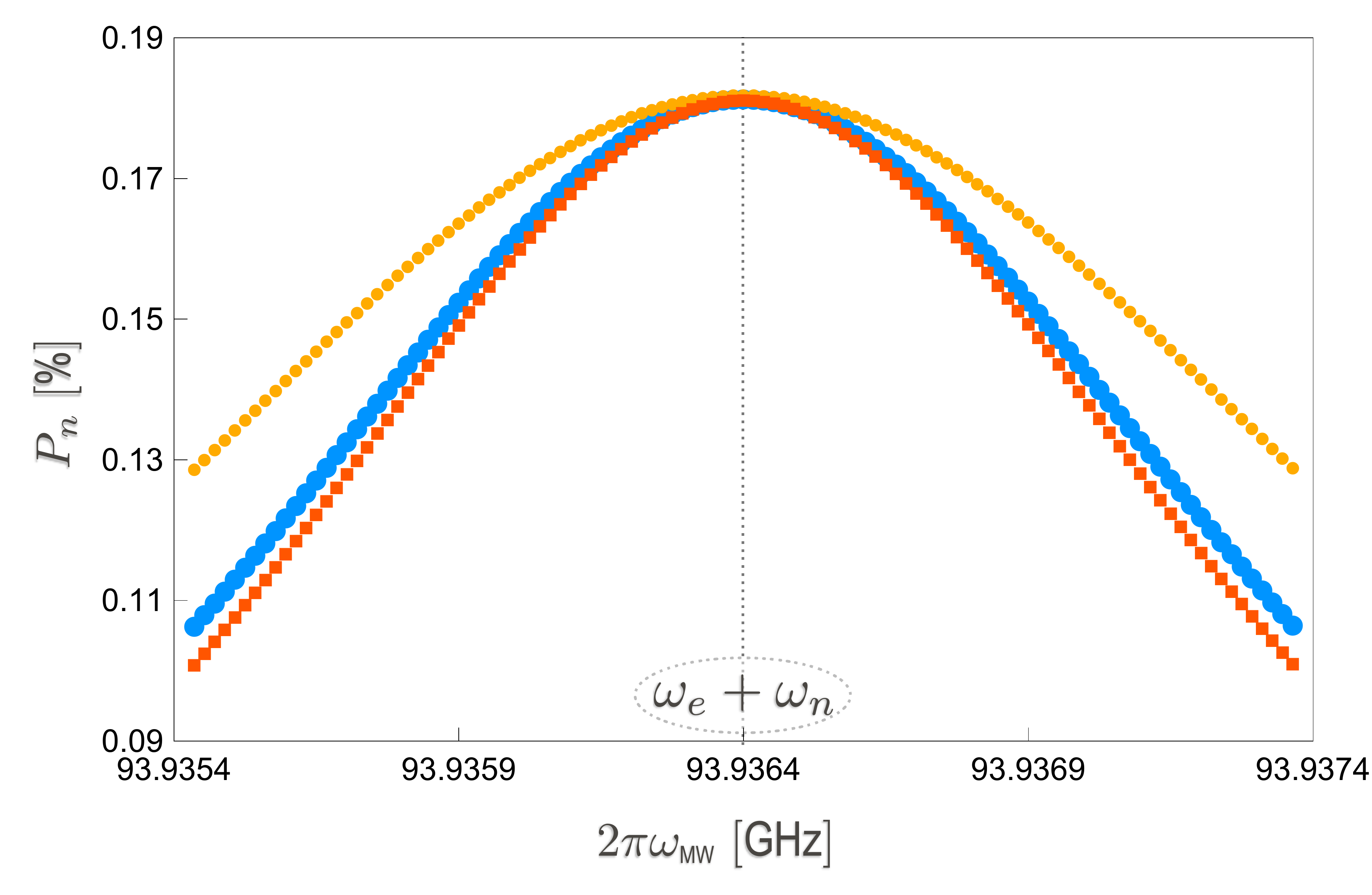} 
\caption{(color online)  DNP profile around the zero-quantum transition within the Hilbert formalism for a hyperfine interacting strength of $B = 320 \times2\pi$kHz. We compare the Zeeman (light blue) and eigenstate-based (yellow) approaches. In red squares, we show the results for the Zeeman-based approach projected on the exact eigenstates, that almost overlaps with the only Zeeman approach. }
\label{HilbertZeemanInt}
\end{figure}
\end{center}

We now explore how the hyperfine interactions affect the performance of the Hilbert approach with respect to the exact Liouville treatment. In Fig~\ref{HilbertPlot} (left and right) we compare the exact Liouville treatment in solid and dashed lines for the Zeeman and eigenstate-based approaches, respectively. Again, the values $B = 40 \times2\pi$kHz and $B = 160\times2\pi$kHz are considered. We observe that 
\begin{itemize}
\item[-]Within the Zeeman-based approach (left) the two treatments slightly differ for the weak hyperfines.  
\item[-]For the eigenstate-based approach (right) Liouville and Hilbert treatments give perfectly matching results.
\end{itemize}
In both cases, as one increases the hyperfine interactions strength the two treatments give more and more different results. Note that the loss of resolution in the resonance peak as we increase the hyperfine strength is more remarkable in the Hilbert scheme than in the Liouville one. 

Summarizing, we observe that when the hyperfine interactions are weak, the Zeeman and the eigenstate-based approaches give similar results. However, in general the Zeeman-based approach is only accurate for small values of $B$: it underestimates the nuclear polarization and the effect becomes more and more evident increasing the value of $B$. In the next section, we discuss further the origin of this discrepancy.

\subsection{The role of leakage \label{sec:leak}}
To investigate the difference between the Zeeman and the eigenstate-based approaches, we now consider a much larger hyperfine strength $B = 320\times2\pi$kHz. The results are shown in Fig.~\ref{HilbertZeemanInt} and, similarly to what we observed in Fig.~\ref{HilbertPlot}, the Zeeman approach (blue circles) leads to a smaller value of the nuclear polarization compared with the eigenstate-based approach (yellow dots). 
However, it is hard to understand why the two differ, as the two polarizations are the result of two separate steps: i) the Liouville description, obtained considering $\HH{0} = \Hs$ (eigenstate-based approach) or  $\HH{0}=\HH{Z}$ (Zeeman-based approach); ii) the Hilbert scheme was obtained projecting the density matrix $\rho$ onto its diagonal in the basis of eigenstates of $\HH{0}$. 

Instead of following ii), it is possible to achieve a quantitative understanding of the difference between the two Liouville formulations, if for both of them, we perform a Hilbert projection onto the same basis, i.e. the eigenstates of $\Hs$. 
Of course, for the eigenstate-based approach, this projection coincides with the already-considered Hilbert scheme, plotted in yellow in Fig.~\ref{HilbertZeemanInt}. However, for the Zeeman-based approach, we obtain a transition matrix, still in the basis of eigenstates of $\Hs$ (red squares in Fig.~\ref{HilbertZeemanInt}). Now, the two transition matrices can be compared and remarkably, the difference between the two is pinned down as an extra transition term 
\begin{equation}
W_{\ket{\tilde{0}}\to \ket{\tilde{1}}}= \frac{\sin^2 2\varphi}{4T_{2,n}}\;,
\end{equation}
it is clear that this additional transition suppresses the nuclear polarization; moreover in the limit of small $B$ coincides with the one introduced in Eq.~\eqref{rateLeak}. 
So, in practice, the choice of the Zeeman basis in deriving the Liouville formulation already induces a fictitious leakage term, which suppresses the polarization when compared with the eigenstate-based approach.

\section{Conclusions}
In this paper, we have studied one of the simplest DNP mechanisms, the solid effect in a two-spins system, within two different approaches.
On the one hand, the Zeeman-based approach treats perturbatively the hyperfine interaction and considers jumps between completely polarized eigenstates. On the other hand, the eigenstate-based approach treats exactly the hyperfine interactions, and jumps occur between eigenstates which present mixing for the nuclear spin. The two schemes give very similar results when the hyperfine interactions are weak. The Zeeman approach could be more convenient for many-body systems, as it does not require the exact diagonalization of the Hamiltonian of size $2^{N_e+N_n}$. However, the validity of this method requires weak interactions between spins. Indeed, our results show that in presence of strong hyperfine interactions the Zeeman and eigenstate-based schemes disagree and the hyperfine interactions must then be treated exactly. As discussed in Sec.~\ref{sec:leak}, the origin of the two different behaviors is not in the precise form of the eigenstates, but rather in the presence of parasite leakage transitions, induced by the perturbative treatment of the hyperfine interactions in the Zeeman approach.

\vspace{0.2cm}
It is a pleasure for us to acknowledge our sponsor, the ANR-16-CE30-0023-01(THERMOLOC).

\bibliography{dnp}

\pagebreak

\pagebreak

\onecolumngrid
\begin{center}
{\Large Supporting information \\ 
\titleinfo
}
\end{center}
In the supporting information we provide the technical details of our computations. In App. \ref{Lattice_coup} we develop the formalism to treat a system weakly coupled to a lattice at a given temperature. In App.~\ref{RWA} we provide the details on the rotating wave approximation to treat time-dependent periodical Hamiltonians. In App.~\ref{SW_PT} we develop the perturbation theory of Schrieffer-Wolf type that allows us to systematically compute the transition rates between eigenstates of the Hamiltonian. Finally, in App.~\ref{RateTr} we summarize all the transition rates between those eigenstates in the different scenarios introduced in the main text.

\section{Formal treatment of the coupling between system and lattice \label{Lattice_coup}}

Consider a system described by the Hamiltonian $\Hs$. The system is not isolated but in contact with a larger system described by $\HH{L}$, that we will call the \emph{lattice}. The latter is considered to be a very large reservoir at equilibrium at the external temperature $\beta^{-1}$. The coupling between both system and lattice is considered to be weak so that the state of the system will not affect that of the lattice. We first deal with the eigenstate-based approach for which $H_0=\Hs$ and the total Hamiltonian reads
\begin{equation}
\HH{\text{tot}}=\Hs+\HH{L}+\lambda\HH{S-L}\;, \text{ with }\; \lambda\HH{S-L}= \sum_{\substack{\alpha=x,y,z\\j}}  \lambda_j \hat{O}^i_\alpha\phi^i_\alpha \;,
\end{equation} 
where $j$ labels if we are referring to the electron or nuclear spin, $\hat{O}^j_\alpha$ is the respective spin operator in the direction $\alpha$ and $\hat{\phi}_\alpha^j$ represents the lattice modes that linearly couple (with coupling constant $\lambda_j$) to the spin $\hat{O}^j_\alpha$. Within the assumptions of the \emph{Born-Markov} approximation detailed in the main text, one can perform a perturbative theory in the coupling $\lambda$ and find~\cite{petruccione2002theory} an integro-differential equation for the reduced density matrix of the spins system $\rhos$ provided that $\rho_{\text{tot}}=\rhos\otimes\rhol$ that reads:
\begin{equation}
\label{int_tev}
\frac{d\rhos(t)}{dt}=-\int_0^\infty ds\, \underset{\text{ lattice}}{\Tr} 
\left[\lambda \HH{S-L}(t),\left[\lambda\HH{S-L}(t-s),\rhos(t)\otimes\rhol \right]\right] \;.
\end{equation}
In order to solve~\eqref{int_tev} one would like to decompose the interaction Hamiltonian into eigenoperators of the Hamiltonian of the spins system. If we note the eigenstates of the unperturbed Hamiltonian as $\ket{n},\ket{m}$ with respective energies $\epsilon_n,\epsilon_m$, we can define the operators
\begin{equation}
\hat{O}_\omega=\sum_{\substack{n,m/\\ \epsilon_n-\epsilon_m=\hbar\omega}}\ket{n}\bra{n}\hat{O}\ket{m}\bra{m}\;,
\end{equation}
where the sum runs over all pairs of eigenstates provided that they differ from an energy $\hbar\omega$. From the definition of these projected operators, we get that
\begin{equation}
\hat{O}^\dagger_\omega = \hat{O}_{-\omega}
\end{equation}
Additionally, one can verify that their time-evolution reads:
\begin{equation}
\hat{O}_\omega (t)= e^{i\Hs t}\,\hat{O}_\omega \,e^{-i\Hs t/\hbar} = e^{-i\omega t}\, \hat{O}_\omega\;.
\end{equation}
If we now sum over all the energy gaps, we obtain the original operator:
\begin{equation}
\sum_\omega \hat{O}_\omega = \hat{O}\;
\end{equation}
As a result, we get that the Hamiltonian of the coupling between lattice and system reads:
\begin{equation}
\lambda\HH{S-L}= \sum_{\substack{\alpha=x,y,z\\j,\omega}}  \lambda_j \hat{O}^j_{\alpha,\omega}\phi^j_\alpha \;\Longrightarrow 
\lambda\HH{S-L}(t)= \sum_{\substack{\alpha=x,y,z\\j,\omega}}  \lambda_j e^{-i\omega t}\hat{O}^j_{\alpha,\omega}\phi^j_\alpha \;.
\end{equation}
Finally, we can substitute these definition in Eq.~\eqref{int_tev} and obtain after some algebra that
\begin{equation}
\label{tev_four}
\frac{d\rhos(t)}{dt}=\sum_{\substack{\alpha,\beta=x,y,z\\j,\omega,\omega'}}e^{i(\omega'-\omega)t}\, \Gamma_{\alpha\beta}^j(\omega) \left[
\hat{O}_{\beta,\omega}^j \rhos(t) \hat{O}_{\alpha,\omega'}^{j\dagger } - \hat{O}_{\alpha,\omega'}^{j\dagger }\hat{O}_{\beta,\omega}^j\rhos(t)
\right] + h.c.
\end{equation}
Here, $h.c.$ stands for Hermitian conjugated, and if we assume spatial and time homogeneity then the correlation functions of the lattice:
\begin{equation}
\Gamma_{\alpha\beta}^j(\omega) = \delta_{\alpha\beta}\int_0^\infty ds \,e^{i\omega s}\,\underset{\text{ lattice}}{\Tr}\left[\rhol\, \hat{\phi}_\alpha^j(s) \hat{\phi}_\alpha^j(0) \right]\;.
\end{equation}
Finally, one can perform the so called \emph{secular} approximation, which is an analogous to the \emph{rotating wave} approximation in order to remove the time-dependence in Eq.~\eqref{tev_four}. Retaining only the terms with $\omega'=\omega$ is justified if the correlation times of the spins system are much larger than the characteristic times of the correlation functions of the lattice. In that case, for time variations where we appreciate a change of $\rhos$ the exponential in~\eqref{tev_four} oscillates very rapidly, so they are averaged out. 
At the end, we obtain that the full time-evolution of the reduced density matrix of the spins system (without taking into account the microwaves yet) reads
\begin{equation}
\label{eqEvol}
\frac{d\rhos}{dt} = -\frac{ i}{\hbar} \left[\Hs,\rhos\right]+ \mathcal{L}[\rhos]\;,
\end{equation}
with $\mathcal{L}$ the Lindblad super-operator that acts on the density matrix as follows:
\begin{equation}
\label{lindApp}
\mathcal{L}[\bullet] =\sum_{\substack{\alpha=x,y,z\\j,\omega}}
 J^j(\omega)
 \left( \hat{O}_{\alpha,\omega}^j\bullet \hat{O}_{\alpha,\omega}^{j\dagger}-\frac{1}{2}\left\{\hat{O}_{\alpha,\omega}^{j\dagger} \hat{O}_{\alpha,\omega}^j,\bullet \right\}\right)\;,
\end{equation}
and all the role of the lattice is encoded in the spectral function
\begin{equation}
J^j(\omega)=\sum_{\alpha=x,y,z}\int_{-\infty}^{\infty} ds\; e^{i\omega s}\underset{\text{ lattice}}{\Tr}\left[\rhol\,\hat{\phi}_\alpha^j(s) \hat{\phi}_\alpha^j(0) \right]\;.
\end{equation}

In the main text, we have substituted the sum over $\alpha$ and $j$ in Eq.~\eqref{lindApp} by the sum over $\hat{O}\in\mathcal{O}=\{\hat{S}_x,\hat{S}_y,\hat{S}_z,\hat{I}_x,\hat{I}_y,\hat{I}_z\}$. Note that the definition of the Lindblad super-operator strongly depends on the approach that we follow, as the operators $\hat{O}_{\alpha,\omega}$ that enter in equation~\eqref{eqEvol} are projected precisely onto the eigenstates of the unperturbed Hamiltonian $\hat{H}_0$. In particular, when one implements the Zeeman approach, the evolution equation is still given by~\eqref{eqEvol}, but the operators $\hat{O}_{\alpha,\omega}$ are projected onto the Zeeman basis.

\section{The rotating frame and the rotating wave approximation}\label{RWA}
The \emph{rotating wave} approximation is one of the options to which we can turn in order to treat time-periodic Hamiltonians like the microwave drive in~\eqref{MWham}. To do so, we consider a new frame of reference that is rotating around the $z$-axis at the same frequency of the periodical Hamiltonian, $\omw$. This can be expressed by writing a new state
\begin{equation}
\ket{\psi^{\text{(r)}}} =e^{i \hat{S}_z\omw t/\hbar} \ket{\psi}\;,
\end{equation}
which in the density matrix language reads:
\begin{equation}
\rhor (t)= \Ut\rho(t) \,\Udt\;,
\end{equation}
introducing the operator $\Ut=e^{i \hat{S}_z\omw t/\hbar}$. If we want to study the time-evolution of this recently introduced density matrix, we simply apply the chain rule as
\begin{equation}
\frac{d\rhos^{\text{(r)}}}{dt}=\frac{d}{dt}\left[\Ut\rho(t) \,\Udt\right] =\Ut \frac{d\rhos}{dt}  \,\Udt + i\omw\left[ \hat{S}_z,\rhos^{\text{(r)}} \right]\;.
\end{equation}
To compute this, we introduce the time-evolution of the density matrix from the main text in~\eqref{time_ev}. We thus obtain that
\begin{align}
\frac{d\rhos^{\text{(r)}}}{dt} = &
-\frac{ i}{\hbar} \Ut \left[\HH{Z} + \Hhf + \Hmw,\rhos\right] \,\Udt  
&+& \;\;\;\;\; i\omw\left[ \hat{S}_z,\rhos^{\text{(r)}} \right] 
&+& \;\;\;\;\; \Ut \mathcal{L}[\rhos] \,\Udt  \nonumber \\
=& 
-\frac{ i}{\hbar} \left[\HH{Z} + \Hhf -\hbar\omw\hat{S}_z ,\rhos^{\text{(r)}} \right]  &-&
\;\;\;\;\; \frac{i}{\hbar}\Ut \left[\Hmw,\rhos\right] \,\Udt &+&
\;\;\;\;\; \Ut \mathcal{L}[\rhos] \,\Udt \nonumber\\
=&
 -\frac{ i}{\hbar} \left[\HH{Z}^{\text{(r)}} + \Hhf  ,\rhos^{\text{(r)}} \right]  &-&
\;\;\;\;\;  \frac{i}{\hbar}\left[\Hmw^{\text{(r)}},\rhos^{\text{(r)}} \right]  &+&
\;\;\;\;\; \Ut \mathcal{L}[\rhos] \,\Udt\;.
\end{align}
Here, in the first step, we took into account the fact that both $\HH{Z}$ and $\HH{hf}$ commute with $\Ut$. In the second step, we took into account the fact that $\Ut\Udt= \mathds{1}$ and redefine the microwave Hamiltonian in the rotating frame as:
\begin{equation}
\Hmw^{\text{(r)}} = \Ut \Hmw \,\Udt\approx \hbar\omega_1\hat{S}_x.
\end{equation}
Note that we have neglected the terms that oscillate with a frequency $2\omw$, as they are soon averaged out. Finally, we want to see the action of the rotating frame on the Lindblad super-operator. 
\begin{equation}
\Ut \mathcal{L}[\rhos] \,\Udt
\end{equation}
It is clear that the only action of the rotating frame for the subset of super-operators acting on the nuclear spin (i.e. $\mathcal{L}_{\hat{O}^n_{\alpha,\omega}}$) has as only role of projecting the static density matrix into the rotating frame, as the operators commute:
\begin{equation}
\Ut \mathcal{L}_{\hat{O}^n_{\alpha,\omega}}[\rhos] \Udt = \mathcal{L}_{\hat{O}^n_{\alpha,\omega}}[\rhos^{\text{(r)}}] \;,
\end{equation}
where the superscript $n$ stands for \emph{nuclear spin}. Additionally,  we will prove that this is also the case for the super-operators associated to the electron spin labeled with the superscript $e$. One can see that:
\begin{equation}
\Ut\; \hat{O}_{\alpha,\pm\omega}^e\; \Udt = 
\sum_{\substack{n,m/\\ \epsilon_n-\epsilon_m=\hbar\omega}} 
e^{\pm i (s_{z,n}-s_{zm}) \omw t}
\ket{n}\bra{n} \hat{O}_{\alpha}^e \ket{m}\bra{m} \;.
\end{equation}
This translates to the fact that for any operator $\hat{O}_{\alpha,\omega}^j$ we get
\begin{equation}
\Ut \mathcal{L}_{\hat{O}_{\alpha,\omega}^j} \left[\rhos \right] \Udt = \mathcal{L}_{\hat{O}_{\alpha,\omega}^j} \left[\rhor \right] \;,
\end{equation}
as in every term of the definition of the Lindbladian~\eqref{lind} we find the product $\hat{O}_{\alpha,\omega}^j \hat{O}_{\alpha,\omega}^{j\dagger}$. 


\section{Schrieffer-Wolf perturbation theory for non-Hermitian operators \label{SW_PT}}
In this appendix we develop the perturbation theory based on the Schrieffer-Wolf transformation. 
To be general we assume to have a spin system composed by $N_n$ nuclear spins and $N_e$ electron spins.
Consider the Liouville equation for the density matrix, which takes the form
\begin{equation}
\label{linearsyst}
\dot{\rho}=L\rho=(L_0+\epsilon V)[\rho]\;.
\end{equation}
From the mathematical point of view, this is a linear differential equation and $\rho$ is a vector with $N = 2^{2(N_e + N_n)}$ components. 
As discussed in the text, $L_0$ preserves the diagonal part of the density matrix $\rho$, which translates into 
\begin{equation}
L_0 [\ket{n} \bra{n}] = 0
\end{equation}
where $\ket{n}$ is an eigenstate of $\HH{0}$, which as discussed in the text can be $\HH{0} = \Hs$ (eigenstate-based approach) or $\HH{0} = \HH{Z}$ (Zeeman-based approach). 
It is clear that for $\epsilon = 0$, each projector $\ket{n} \bra{n}$ would be a stable stationary state. For $\epsilon \neq 0$ but small, we can derive an effective dynamics restricted to the eigenspace of $L_0$ with $0$-eigenvalue
\begin{equation}
\mathcal{G}_0 = \Ker(L_0 - \lambda_0 \mathbf{1}) \;.
\end{equation}
where we introduced $\lambda_0 = 0$, to keep the treatment general to any eigenspace of $L_0$. 
Note that $L_0$ has a large degeneracy, as $\dim \mathcal{G}_0 = 2^{N_e + N_n}$. Moreover, the validity of the expansion is quantified by
\begin{equation}
 \Delta = \min_{\substack{\lambda \in \sigma(L_0) \\ \lambda \neq \lambda_0}} | \lambda - \lambda_0| \gg \epsilon || V || 
\end{equation}
and $\sigma(L_0)$ indicates the spectrum of $L_0$: $L_0 \rho_\lambda = \lambda \rho_\lambda $. In other words, the perturbation $\epsilon V$ is assumed to be too small to generate transition outside of the subspace $\mathcal{G}_0$; nevertheless, it can induce \textit{virtual} transitions, which by going outside and back inside $\mathcal{G}_0$, can generate an effective dynamics within the subspace. 

To quantitatively compute the dynamics within $\mathcal{G}_0$, we use an analogous of the Schrieffer-Wolf transformation. 
The idea behind this method is to consider a linear transformation $\rho \to U \rho$. Under this transformation, the matrix $L$ in Eq.~\eqref{linearsyst} is transformed as $U L U^{-1}$. We look for a transformation $U$ such that the subspace $\mathcal{G}_0$ remains decoupled from the rest. Of course, if we were able to find the transformation $U$ which completely diagonalizes $L$, we would have decoupled $\mathcal{G}_0$ from all the other subspaces. But, the diagonalization of $L$ is the hard problem that we want to avoid, so we settle for the simpler requirement of decoupling $\mathcal{G}_0$ from the other subspaces, order by order in $\epsilon$. 

Let's indicate with $P$ the projector on the subspace $\mathcal{G}_0$ and $Q = \mathbf{1}-P$. We also introduce the projectors $P_\lambda$ onto all the other eigenspaces $\mathcal{G}_\lambda$ of $L$, such that $P_\lambda \mathcal{G}_{\lambda'} =  \delta_{\lambda,\lambda'} \mathcal{G}_{\lambda'}$ and
\begin{align}
 &\sum_{\lambda \in \sigma(L_0)} P_\lambda = P + \psum_\lambda P_\lambda = P + Q = \mathbf{1} \;.
 \end{align}
where we use the notation $\psum$ to indicate the sum over all the eigenvalues but $\lambda = \lambda_0 = 0$.

It is useful to set $U = e^{i S}$ and  we then demand that 
\begin{equation}
\label{swcond}
Q e^{i S} (L_0 + \epsilon V) e^{-i S} P = 0\;, \qquad P e^{i S} (L_0 + \epsilon V) e^{-i S} Q = 0\;,
\end{equation}
The effective operator $\Leff$ in the subspace $\mathcal{G}_0$ can then be written as
\begin{equation}
\label{AnsSolution}
\Leff = P e^{i S} L e^{-i S} P\;.
\end{equation}
This equation can be solved perturbatively in $V$, by writing $S = S^{(0)} + \epsilon S^{(1)} + \ldots$.
At order zero, we trivially obtain $S^{(0)} = 0$ as $Q L_0 P = 0$. In the following we will derive the subsequent orders.

At first order, we get the conditions that $S^{(1)}$ has to verify:
\begin{equation}
\label{firstorder}
 Q\left(i [S^{(1)}, L_0] + V\right)P = 0 \;,  \quad P\left(i [S^{(1)}, L_0] + V\right)Q = 0\;.
\end{equation}
One can introduce an Ansatz for $ S^{(1)} $ with the possible combinations of $P$ and $P_{\lambda}$ connected once by $V$:
\begin{equation}
 S^{(1)} = \psum_\lambda a_\lambda P_\lambda V P + b_\lambda P V P_\lambda
\end{equation}
and substitute it in Eq.~\eqref{firstorder} and using that $PQ = PP_\lambda = 0$ for $\lambda \neq \lambda_0$, it is easy
to get the value of $a_\lambda$ and $b_\lambda$. We finally obtain:
\begin{equation}
\label{S1sol}
 S^{(1)} = \psum_\lambda \frac{i}{\lambda_0 - \lambda} (P_\lambda V P - P V P_\lambda)\;.
\end{equation}
In general the matrices $S^{(j)}$ are ``off-diagonal'', in the sense that they connect the spaces $G_0$ with the rest (and vice-versa).
It means that the matrix $S^{(1)}$ is already enough to obtain the effective operator at second order in $\epsilon$. Expanding the solution in Eq.~\eqref{AnsSolution}, and after some algebra, we obtain the first order contribution to the effective operator:
\begin{equation}
 \label{effectivefirstorderexpl}
  \Leff^{(1)} = 
PVP  +  \psum_{\lambda} \frac{1}{\lambda_0 - \lambda}  P V P_\lambda V P 
\end{equation}
As anticipated in the main text, one has to get to the fourth order in the perturbation theory in order to obtain the solid effect transitions. The procedure is analogous to the one discussed before, and the steps are detailed in the following. At second order, the conditions in Eq.~\eqref{swcond} lead to:
\begin{subequations}
\label{secondorder}
\begin{align}
 Q\left(i[S^{(2)}, L_0] + i[S^{(1)}, V] + S^{(1)} L_0 S^{(1)} - \frac{1}{2} \{ (S^{(1)})^2, L_0 \}\right)P = 0 \;,  \\
 P\left(i[S^{(2)}, L_0] + i[S^{(1)}, V] + S^{(1)} L_0 S^{(1)} - \frac{1}{2} \{ (S^{(1)})^2, L_0 \}\right)Q = 0 \;.
 \end{align}
 \end{subequations}
We easily see that the last two terms in \eqref{firstorder} do not contribute as they do not connect $\mathcal{G}_0$ with the outside. As before, we find the solution using introducing the Ansatz for $S^{(2)}$ (this time with the perturbation $V$ acting twice):
\begin{equation}
 S^{(2)} = \psum_\lambda a_0 P V P V P_\lambda +  a_1 P_\lambda V P V P +  \psum_{\lambda,\lambda'}
 b_0 P V P_{\lambda'} V P_{\lambda} +  b_1 P_\lambda V P_{\lambda'} V P 
\end{equation}
in~\eqref{secondorder}. Again one can obtain the coefficients $a_0,a_1,b_0,b_1$ and after simplifications compute the final expression for $\Leff$ up to third order:
\begin{equation}
 \label{effectivesecondorderexpl}
 \Leff^{(2)} = 
- \psum_{\lambda} \frac{1}{2(\lambda_0 - \lambda)^2} ( PVPVP_\lambda VP + PVP_\lambda VPVP)
 + \psum_{\lambda, \lambda'} \frac{P V P_\lambda V P_{\lambda'} V P}{(\lambda_0 - \lambda) (\lambda_0 - \lambda')}
\end{equation}
The same procedure must be implemented one more time and, at third order, the conditions fixed by Eq.~\eqref{swcond} read:
\begin{subequations}
\begin{align}
 Q\left(
 i[S^{(3)}, L_0] + i[S^{(2)}, V] + S^{(1)} V S^{(1)} - \frac{1}{2} \{ (S^{(1)})^2, V \} - \frac{1}{2} \{ L_0, \{ S^{(1)}, S^{(2)}\}  \}
 + S^{(1)}L_0 S^{(2)}  \right. 
 \nonumber \\
\left. + S^{(2)}L_0 S^{(1)} +\frac{i}{2} ( (S^{(1)})^2 L_0 S^{(1)} - S^{(1)}L_0 (S^{(1)})^2)
-\frac{i}{6} [(S^{(1)})^3, L_0]
 \right)P = 0 \;,  \\
  P\left(
 i[S^{(3)}, L_0] + i[S^{(2)}, V] + S^{(1)} V S^{(1)} - \frac{1}{2} \{ (S^{(1)})^2, V \} - \frac{1}{2} \{ L_0, \{ S^{(1)}, S^{(2)}\}  \}
 + S^{(1)}L_0 S^{(2)}  \right. 
 \nonumber \\
\left. + S^{(2)}L_0 S^{(1)} +\frac{i}{2} ( (S^{(1)})^2 L_0 S^{(1)} - S^{(1)}L_0 (S^{(1)})^2)
-\frac{i}{6} [(S^{(1)})^3, L_0]
 \right)Q = 0 \;.
 \end{align}
 \label{eq3rd}
 \end{subequations}

A new Ansatz up to third order for the form of $S^{(3)}$ is again introduced:
\begin{align}
 S^{(3)} & =  \psum_\lambda a_0 P V P V P V P_\lambda + a_1 P_\lambda V P V P V P  
 +\nonumber \\ &+  \psum_{\lambda,\lambda'} 
 b_0 P V P V P_{\lambda'} V P_{\lambda} +  b_1 P V P_{\lambda'} V P V P_{\lambda}   +  b_2  P_{\lambda'} V P V P_{\lambda} V P +  b_3 P_{\lambda'} V P_{\lambda} V P V P +\nonumber \\
  & +  \psum_{\lambda,\lambda',\lambda''} 
   c_0 P V P_{\lambda''} V P_{\lambda'} V P_{\lambda} +  c_1 P V P_{\lambda''} V P_{\lambda'} V P_{\lambda} 
\end{align}
and its coefficients can be determined by imposing \eqref{eq3rd}. After several simplifications, one finally obtains
 \begin{align}
  \label{effectivethirdorderexpl}
 \Leff^{(3)} = 
\psum_{\lambda} \frac{1}{2(\lambda_0 - \lambda)^3} ( PVPVPVP_\lambda VP + PVP_\lambda VPVPVP)
 + \psum_{\lambda, \lambda',\lambda''} \frac{P V P_\lambda V P_{\lambda'} VP_{\lambda''} V P}{(\lambda_0 - \lambda) (\lambda_0 - \lambda')(\lambda_0 - \lambda'')}\nonumber\\
 +\psum_{\lambda,\lambda'} \frac{-2\lambda_0+\lambda+\lambda'}{2(\lambda_0-\lambda)^2(\lambda_0-\lambda')^2} 
 ( PVPVP_\lambda VP_{\lambda'} VP + PVP_\lambda VP_{\lambda'}VPVP + PVP_\lambda VPVP_{\lambda'}VP)
\end{align}
The final effective solution for the operator $L$ is up to fourth order:
\begin{equation}
\Leff=\Leff^{(1)}+\Leff^{(2)}+\Leff^{(3)}\;.
\end{equation}
As it is clear, at the fourth order, one obtains a large number of terms. For the sake of simplicity, we have kept for each transition between a pair of eigenstates $\ket{n}$ and $\ket{m}$ the lowest order at which it does not vanish. For example, the lattice-induced transitions can be simply obtained with the first order contribution $PVP$. The solid effect transitions, on the contrary, must be obtained using Eq.~\eqref{effectivethirdorderexpl}. 

\pagebreak
\section{Transition rates\label{RateTr}}
In this appendix we summarize the different rates obtained using the lowest order of the perturbation theory computed in the previous appendix. 
\begin{itemize}
\item \emph{Transitions for the eigenstate-based approach}: 

The terms come straightforwardly from the first and second-order of the perturbation theory in Eqs.~\eqref{rates_T1} and~\eqref{2ndOrderRate}, respectively. It is thanks to the mixing between the nuclear eigenstates in Eq.~\eqref{StatesNucl} that the solid-effect transitions have a non-vanishing contribution. 

\begin{subequations}
\label{ratesEigenstates}
\begin{align}
W_{\ket{\tilde{0}}\to \ket{\tilde{1}}}=\;&W_{\ket{\tilde{2}}\to \ket{\tilde{3}}}=\frac{h(\Omega_n)}{4T_{1,n}} (1+\cos^2 2\varphi )
\; \\
W_{\ket{\tilde{1}}\to \ket{\tilde{0}}}=\;&W_{\ket{\tilde{3}}\to \ket{\tilde{2}}}=\frac{h(-\Omega_n)}{4T_{1,n}}(1+\cos^2 2\varphi )
\;
\\
W_{\ket{\tilde{0}}\to \ket{\tilde{2}}}=\;&W_{\ket{\tilde{1}}\to \ket{\tilde{3}}}=\frac{h(\omega_e)}{2T_{1,e}}\cos^2 2\varphi
+\frac{ \omega_1^2 T_{2e} \cos^2\varphi/2}{1 + 
 T_{2e}^2 ( \omega_e  - \omw)^2}
\; ,
\\
W_{\ket{\tilde{2}}\to \ket{\tilde{0}}}=\;&W_{\ket{\tilde{3}}\to \ket{\tilde{1}}}=\frac{h(-\omega_e)}{2T_{1,e}} \cos^2 2\varphi
+\frac{ \omega_1^2 T_{2e} \cos^2\varphi/2}{1 + 
 T_{2e}^2 ( \omega_e  - \omw)^2}\;,
\\
&W_{\ket{\tilde{0}}\to \ket{\tilde{3}}}
=
\frac{h(\omega_e+\Omega_n)}{2T_{1,e}} \sin^2 2 \varphi + \frac{ \omega_1^2 T_{2e} \sin^2 2 \varphi/2}{1 + 
 T_{2e}^2 ( \omega_e + \Omega_n - \omw)^2}
\; , 
\\
&W_{\ket{\tilde{3}}\to \ket{\tilde{0}}} = 
\frac{h(-\omega_e-\Omega_n)}{2T_{1,e}} \sin^2 2 \varphi+ \frac{ \omega_1^2 T_{2e} \sin^2 2 \varphi/2}{1 + 
 T_{2e}^2 ( \omega_e + \Omega_n - \omw)^2}
\; , 
\\
&W_{\ket{\tilde{1}}\to \ket{\tilde{2}}}
=
\frac{h(\omega_e-\Omega_n)}{2T_{1,e}} \sin^2 2 \varphi + \frac{ \omega_1^2 T_{2e} \sin^2 2 \varphi/2}{1 + 
 T_{2e}^2 ( \omega_e - \Omega_n - \omw)^2}
\; , 
\\
&W_{\ket{\tilde{2}}\to \ket{\tilde{1}}} = 
\frac{h(-\omega_e+\Omega_n)}{2T_{1,e}} \sin^2 2 \varphi + \frac{ \omega_1^2 T_{2e} \sin^2 2 \varphi/2}{1 + 
 T_{2e}^2 ( \omega_e - \Omega_n - \omw)^2}
\; , 
\end{align}
\end{subequations}

\item \emph{Transitions for the Zeeman approach}: 

All terms, except for the solid effect double-quantum and zero-quantum transitions  ($\ket{3}\to\ket{0},\ket{0}\to\ket{3},\ket{1}\to\ket{2},\ket{2}\to\ket{1}$), can be obtained as in the previous case using the first and second order in the perturbation theory of Eqs.~\eqref{rates_T1} and~\eqref{2ndOrderRate}. In contrast, the second order contribution for the solid-effect transitions vanishes due to the orthogonality of electron and nuclear eigenstates. It's computation requires to go up to the fourth order of the Schrieffer-Wolf perturbation theory developed in appendix~\ref{SW_PT}, that we find in Eq.~\eqref{effectivethirdorderexpl}.

\begin{subequations}
\begin{align}
W_{\ket{0}\to \ket{1}}&=W_{\ket{2}\to \ket{3}}=\frac{h(\omega_n)}{2T_{1,n}} + \frac{A^2 T_{2n}/8}{1+T_{2n}^2 \omega_{n}^2}
\; \\
W_{\ket{1}\to \ket{0}}&=W_{\ket{3}\to \ket{2}}=\frac{h(-\omega_n)}{2T_{1,n}} + \frac{A^2 T_{2n}/8}{1+T_{2n}^2 \omega_{n}^2}
\; 
\\
W_{\ket{0}\to \ket{2}}&=W_{\ket{1}\to \ket{3}}=\frac{h(\omega_e)}{2T_{1,e}}
+\frac{ \omega_1^2 T_{2e}}{1 + 
 T_{2e}^2 ( \omega_e  - \omw)^2}\; ,
\\
W_{\ket{2}\to \ket{0}}&=W_{\ket{3}\to \ket{1}}=\frac{h(-\omega_e)}{2T_{1,e}}
+\frac{ \omega_1^2 T_{2e}}{1 + 
 T_{2e}^2 ( \omega_e  - \omw)^2} \;,
\\
W_{\ket{0}\to \ket{3}}& = W_{\ket{3}\to \ket{0}} = 
\frac{A^2T_{2e}\omega_1^2}{16\omega_n} \left(
\frac{1}{1+T_{2e}^2(\omega_e-\omw)^2} + \frac{2}{1+T_{2e}^2(\omega_e+\omega_n-\omw)^2}
\right)
\; , 
\\
W_{\ket{1}\to \ket{2}}& = W_{\ket{2}\to \ket{1}} =
\frac{A^2T_{2e}\omega_1^2}{16\omega_n} \left(
\frac{1}{1+T_{2e}^2(\omega_e-\omw)^2} + \frac{2}{1+T_{2e}^2(\omega_e-\omega_n-\omw)^2}
\right)
\end{align}
\end{subequations}

\pagebreak
\item \emph{Transitions for the Zeeman-based approach - projected on the exact eigenstates}:

Here we perform the Hilbert projection of the Lindbladian in the Zeeman approach (i.e. with the spectral operators in Eq.~\eqref{spectralOp}) onto the basis of the exact eigenstates of the system. The only difference with the transitions in the in fully eigenstate-based approach of Eqs.~\eqref{ratesEigenstates} is the presence of a term analogous to the leakage that induces transitions between the mixed nuclear states.
\begin{subequations}
\begin{align}
W_{\ket{\tilde{0}}\to \ket{\tilde{1}}}=\;&W_{\ket{\tilde{2}}\to \ket{\tilde{3}}}=\frac{h(\Omega_n)}{4T_{1,n}} (1+\cos^2 2\varphi )
 + \frac{\sin^2 2\varphi}{4T_{2,n}}
\; \\
W_{\ket{\tilde{1}}\to \ket{\tilde{0}}}=\;&W_{\ket{\tilde{3}}\to \ket{\tilde{2}}}=\frac{h(-\Omega_n)}{4T_{1,n}}(1+\cos^2 2\varphi )
 + \frac{\sin^2 2\varphi}{4T_{2,n}}
\;
\\
W_{\ket{\tilde{0}}\to \ket{\tilde{2}}}=\;&W_{\ket{\tilde{1}}\to \ket{\tilde{3}}}=\frac{h(\omega_e)}{2T_{1,e}}\cos^2 2\varphi
+\frac{ \omega_1^2 T_{2e} \cos^2\varphi/2}{1 + 
 T_{2e}^2 ( \omega_e  - \omw)^2}
\; ,
\\
W_{\ket{\tilde{2}}\to \ket{\tilde{0}}}=\;&W_{\ket{\tilde{3}}\to \ket{\tilde{1}}}=\frac{h(-\omega_e)}{2T_{1,e}} \cos^2 2\varphi
+\frac{ \omega_1^2 T_{2e} \cos^2\varphi/2}{1 + 
 T_{2e}^2 ( \omega_e  - \omw)^2}\;,
\\
&W_{\ket{\tilde{0}}\to \ket{\tilde{3}}}
=
\frac{h(\omega_e+\Omega_n)}{2T_{1,e}} \sin^2 2 \varphi + \frac{ \omega_1^2 T_{2e} \sin^2 2 \varphi/2}{1 + 
 T_{2e}^2 ( \omega_e + \Omega_n - \omw)^2}
\; , 
\\
&W_{\ket{\tilde{3}}\to \ket{\tilde{0}}} = 
\frac{h(-\omega_e-\Omega_n)}{2T_{1,e}} \sin^2 2 \varphi+ \frac{ \omega_1^2 T_{2e} \sin^2 2 \varphi/2}{1 + 
 T_{2e}^2 ( \omega_e + \Omega_n - \omw)^2}
\; , 
\\
&W_{\ket{\tilde{1}}\to \ket{\tilde{2}}}
=
\frac{h(\omega_e-\Omega_n)}{2T_{1,e}} \sin^2 2 \varphi + \frac{ \omega_1^2 T_{2e} \sin^2 2 \varphi/2}{1 + 
 T_{2e}^2 ( \omega_e - \Omega_n - \omw)^2}
\; , 
\\
&W_{\ket{\tilde{2}}\to \ket{\tilde{1}}} = 
\frac{h(-\omega_e+\Omega_n)}{2T_{1,e}} \sin^2 2 \varphi + \frac{ \omega_1^2 T_{2e} \sin^2 2 \varphi/2}{1 + 
 T_{2e}^2 ( \omega_e - \Omega_n - \omw)^2}
\; , 
\end{align}
\end{subequations}
\end{itemize}

\end{document}